\PassOptionsToPackage{table}{xcolor}
\documentclass[sigconf,nonacm]{acmart}
\usepackage{multirow}
\usepackage{graphicx}
\usepackage{tabularx}
\usepackage{enumitem}
\usepackage{listings}

\AtBeginDocument{%
  }

\setcopyright{none}
\acmDOI{}
\acmISBN{}
\newcommand{\score}[2]{\ensuremath{#1{\pm}#2}}
\newcommand{\bestscore}[2]{\textbf{#1\textpm#2}}
\newcommand{\secondscore}[2]{\ensuremath{\underline{#1{\pm}#2}}}
\definecolor{baselinegreen}{RGB}{0,128,0}

\begin{document}

\title{HyperIntent: Tracking Intent Shifts for Efficient Reasoning on Dynamic Textual Graphs}

\author{\texorpdfstring{%
Yilin Yao\textsuperscript{1},
Bin Lu\textsuperscript{1\textdagger},
Xinhui Li\textsuperscript{2},
Meng Jin\textsuperscript{3\textdagger},
Xinbing Wang\textsuperscript{1}%
}{Yilin Yao, Bin Lu, Xinhui Li, Meng Jin, Xinbing Wang}}
\affiliation{%
  \institution{\textsuperscript{1}School of Information Science and Electronic Engineering, Shanghai Jiao Tong University}
  \institution{\textsuperscript{2}School of Computer Science, Shanghai Jiao Tong University}
  \institution{\textsuperscript{3}School of Artificial Intelligence, Shanghai Jiao Tong University}
  \city{Shanghai}
  \country{China}}
\email{\{jolinyao, robinlu1209, lixinhui, jinm, xwang8\}@sjtu.edu.cn}

\renewcommand{\shortauthors}{Yao et al.}

\begin{abstract}
Dynamic web interactions continuously evolve as new edges arrive, offering large language models (LLMs) great potential for reasoning over edge-level temporal semantics.
However, continuously arriving edges are often repetitive, obscuring the key structures that truly reflect temporal changes and limiting the efficiency of LLMs. 
A key insight is that beneath the dense stream of edge arrivals lies a sparse evolution of temporal intent, enabling LLMs to be responsively triggered only when such intent shifts, while otherwise inheriting the previous state. To this end, we propose a hypergraph-based framework named \textbf{\textit{HyperIntent}}, which enhances dynamic graph reasoning by parsing differentiated intents, quantifying their temporal discrepancies, and performing shift-triggered semantic updating. Experimental results on destination node retrieval and future link prediction demonstrate that HyperIntent consistently achieves the best overall performance, with average relative Hits@1 gains of 34.0\% and 25.4\% over vanilla TGAT and DyGFormer, respectively. Meanwhile, HyperIntent reduces LLM token consumption by 54.4\%–95.9\% across six datasets while achieving leading predictive performance. In-depth studies further confirm the necessity of shift triggering for identifying structural and semantic temporal changes. The code will be made publicly available upon completion of the review process.
\end{abstract}


\maketitle
\begingroup
\renewcommand{\thefootnote}{\fnsymbol{footnote}}
\footnotetext[2]{Bin Lu and Meng Jin are the corresponding authors.}
\endgroup

\section{Introduction}

Web platforms continuously generate timestamped, text-rich interactions, such as e-mail communications, event knowledge graphs, and e-commerce reviews and transactions, which are naturally modeled as dynamic textual graphs
~\cite{klimt2004enron,trivedi2017knowevolve,mcauley2015image}. The textual information conveyed by these edges characterizes a sequence of emerging events~\cite{zhang2024dtgb}. 
Recent advances in large language models (LLMs) have enabled richer modeling of edge-level textual semantics in dynamic textual graphs, improving the representations learned by temporal graph neural networks (TGNNs)~\cite{zhang2024dtgb,zhang2025cross,wang2026dygrasp}. However, in real-world web environments, edges arrive continuously~\cite{zhang2023tiger}, and many interactions repeat over time~\cite{poursafaei2022towards,yi2025tgbseq}. Such repetition can obscure the sequential dynamics that reflect structure and semantic evolution~\cite{yi2025tgbseq,xu2026seqfilter}. Meanwhile, applying LLMs to abundant and evolving edge text introduces substantial computational overhead~\cite{wang2026dygrasp,xu2026unlocking,chen2024llmgnn}.

\begin{figure}[t]
    \centering
    \includegraphics[width=0.95\columnwidth]{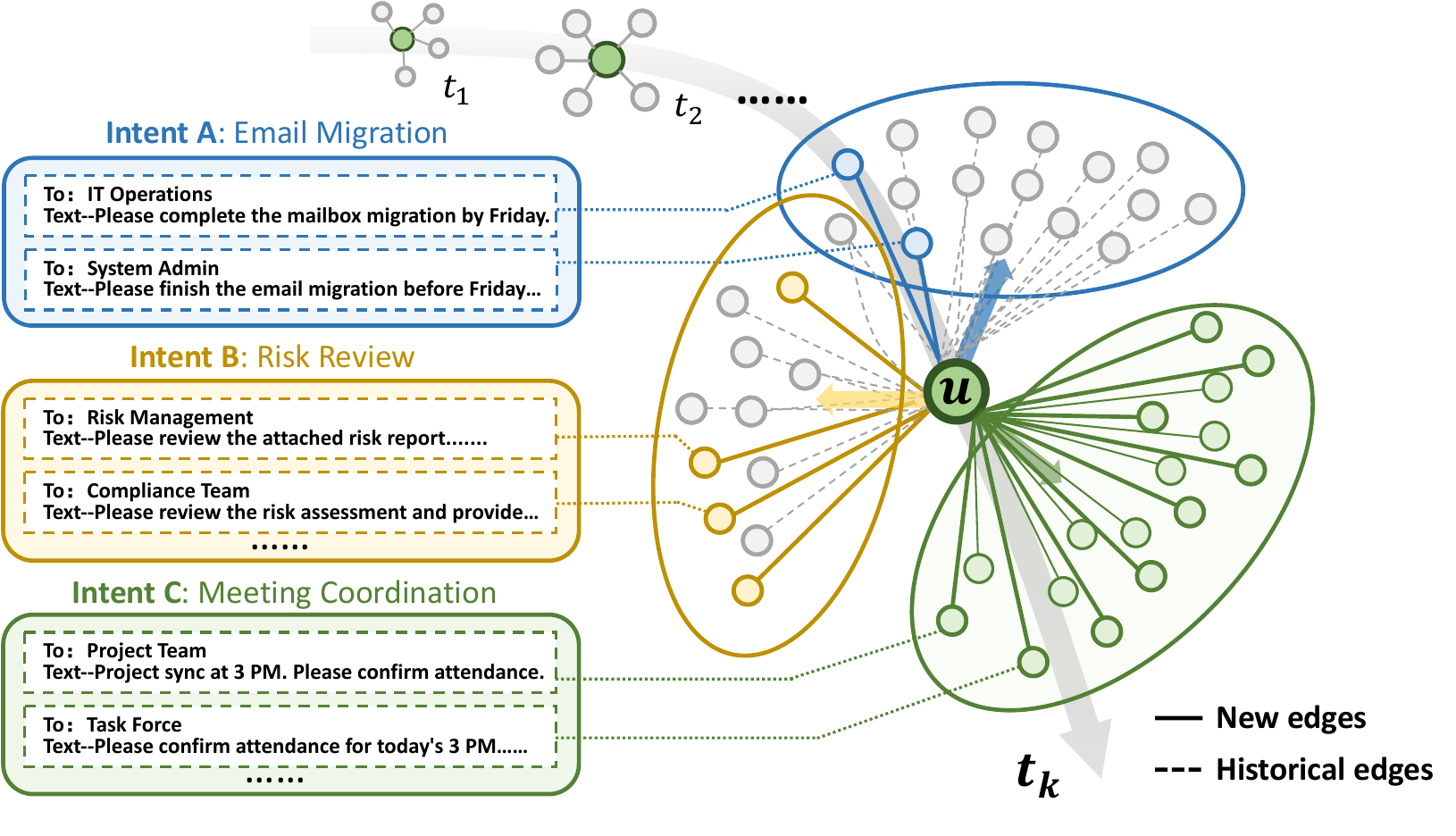}
    \caption{Temporal intent evolution of node \(u\). At \(t_k\), new e-mails (solid edges) form three intents. Email migration and risk review persist from history (dashed edges), whereas meeting coordination emerges as an intent shift.}
    \label{fig:intro}
    \vspace{-3mm}
\end{figure}

Figure~\ref{fig:intro} shows a snapshot of a Web e-mail network at time $t_k$, where the central node $u$ attaches dozens of new edges. The textual content associated with these edges implicitly reveals three underlying intents, i.e., email migration, risk review, and meeting coordination, which effectively summarize the originally complex and redundant interactions. Meanwhile, these intents evolve over time. The emergence of numerous edges related to meeting coordination (highlighted in green) introduces an intent that has not appeared previously, indicating an intent shift of node $u$. In contrast, the other two intents remain consistent with their historical interactions. Such a pattern is prevalent across Web-based dynamic textual graphs. For instance, in academic networks, stable research collaborations coexist with project-driven interdisciplinary partnerships~\cite{ramasco2006social,feng2026interdisciplinary}; in e-commerce networks, routine purchases coexist with promotion-induced transactions~\cite{zhu2024promotions}. These observations reveal that, although dynamic textual graphs may contain massive edge arrivals, the underlying intent evolution is often sparse and localized in time.

Motivated by these observations, our key insight is that LLMs should not be invoked upon every new edge, but responsively triggered by shifted intent. This calls for LLMs to selectively encode edges, an idea recently explored in static graphs~\cite{loveland2026glance,tang2024graphgpt} but substantially more challenging in dynamic settings. (1) Unlike static graphs that quantify edge uncertainty, dynamic graphs require identifying intent shifts, manifested as emerging structural relations or semantic changes. This temporal intent reasoning relies heavily on historical and global contexts~\cite{zhang2025cross,wang2026dygrasp}. (2) Moreover, intent shifts in dynamic graphs typically emerge from collective temporal patterns rather than isolated edges~\cite{dai2025graphpatterns}. Simply concatenating raw edge texts as input to an LLM would only reproduce a fragmented interaction history, making it difficult to capture the coherent intent underlying these interactions~\cite{liu2026ragse,li2024prograph}.

To solve the above issues, we propose \textbf{\textit{HyperIntent}}, a hypergraph-based framework for dynamic LLM-enhanced graph reasoning that transforms the arrival of dense edges into the evolution of sparse intents. HyperIntent organizes similar edge-level interactions into differentiated hyperedges, allowing LLM reasoning on coherent intents. Each intent hyperedge is contrasted with the node’s historical semantic memory and its structural neighborhood to estimate whether either the semantic or the structural discrepancy is significant over time. Only shifted intents trigger the LLM to update the node states, while otherwise directly inherit their historical states, forming an efficient state cache. Extensive experiments on cross-domain datasets, covering destination node retrieval and future link prediction under both transductive and inductive settings, demonstrate the superiority of our HyperIntent. Meanwhile, ablation studies and case analyses both confirm the necessity of shift triggering for identifying temporal changes in structures and semantics. The main contributions are summarized as follows:
\begin{itemize}[leftmargin=1em]
    \item We show that dense streams of edges often correspond to sparse changes in underlying intent, suggesting a responsive framework named HyperIntent that triggers LLMs only upon intent shifts while otherwise inheriting previous states.
    \item HyperIntent leverages hyperedges for differentiated intent parsing and proposes temporal intent discrepancy estimation to identify intent shifts, enabling LLM to encode the intent states and update selectively.
    \item HyperIntent achieves the state-of-the-art performance among all baselines, with relative Hits@1 improvements of 34.0\% over vanilla TGAT and 25.4\% over vanilla DyGFormer, while reducing LLM token consumption by 54.4\%–95.9\% relative to a state-of-the-art LLM-based method across six datasets.
\end{itemize}

\section{Related Work}
\label{sec:related-work}

\subsection{Dynamic Textual Graph Learning }

Temporal graph neural networks learn time-aware representations from timestamped interactions, including recurrent, attention-based, walk-based, and sequence-mixing architectures~\cite{kumar2019jodie,trivedi2019dyrep,xu2020inductive,rossi2020temporal,wang2021inductive,cong2023really,yu2023towards}. DTGB~\cite{zhang2024dtgb} extends this setting to dynamic textual graphs, where evolving structures and free-form text jointly determine future interactions. Its results show that structural temporal signals and static text embeddings do not fully capture temporal evolution.

Motivated by this, recent studies enrich dynamic textual graphs by incorporating LLM-derived semantics into temporal representations. CROSS~\cite{zhang2025cross} and DyGRASP~\cite{wang2026dygrasp} construct semantic contexts from textual neighborhoods and recent-to-global interaction histories, respectively. Although they construct semantic contexts differently, both focus on \emph{what} context to provide to the LLM rather than \emph{when} the LLM should be invoked, so their reasoning cost grows with the volume of arriving interactions. They do not explicitly determine whether newly arriving interactions change a node's structural or semantic state enough to warrant an update. HyperIntent addresses this question by invoking the LLM only for intent hyperedges that reveal such a shift.

\begin{figure*}[!t]
    \centering
    \includegraphics[width=0.85\textwidth]{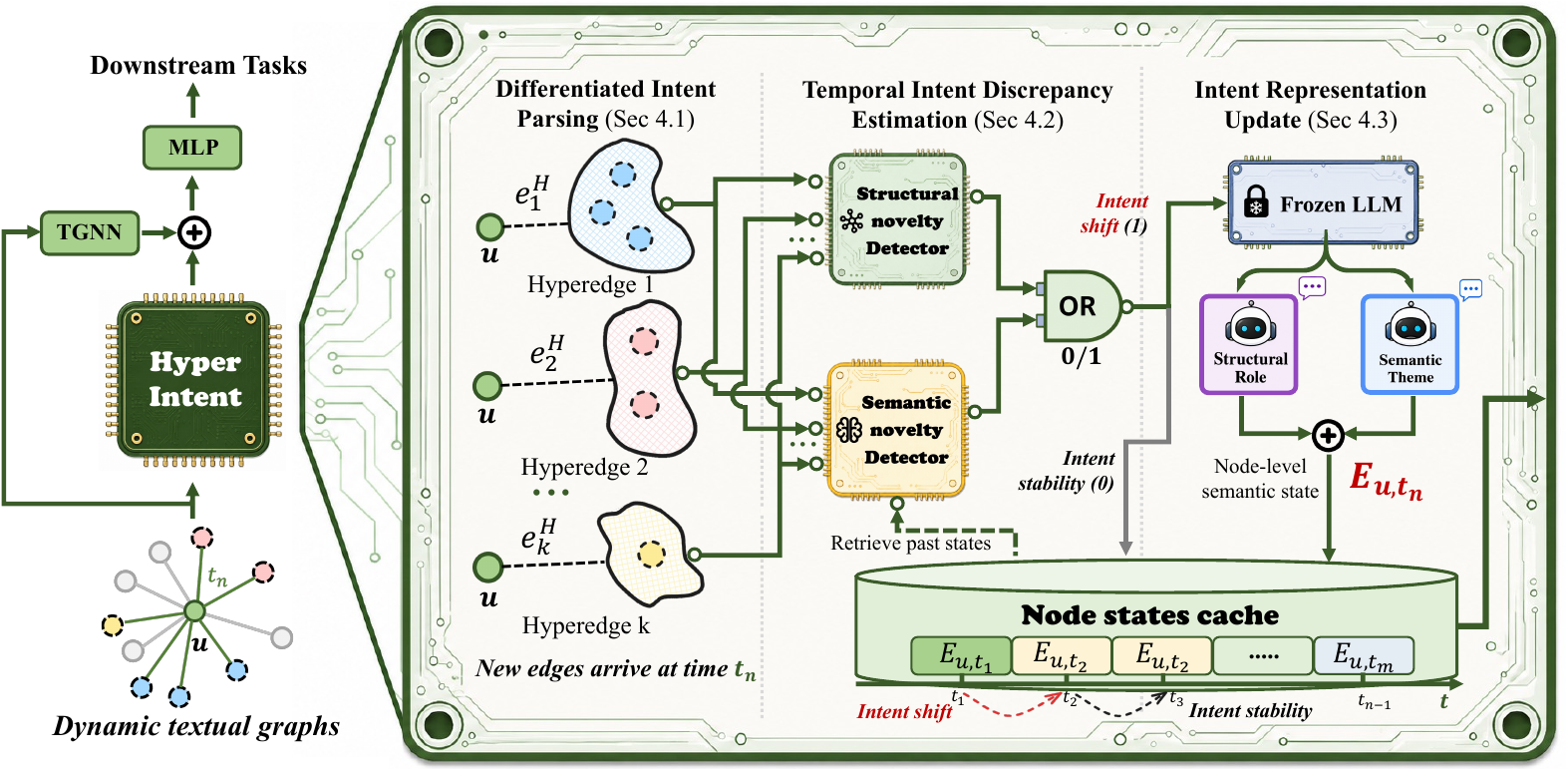}
    \caption{Overview of HyperIntent. Edges arriving at $t_n$ for node $u$ are parsed into intent hyperedges (Sec.~\ref{sec:parsing}) and checked by structural and semantic novelty detectors (Sec.~\ref{sec:discrepancy}). Only shifted intents invoke the frozen LLM to update the cached state $\mathbf{E}_{u,t_n}$, while stable intents inherit the previous state (Sec.~\ref{sec:update}). Cached states are fused with the TGNN representation for prediction.}
    \label{fig:hyperintent-overview}
\end{figure*}

\subsection{Structural Redundancy in Graph Learning}

Structural redundancy is prevalent in temporal graphs~\cite{yuan2024dgib} and often arises from repeated interactions. EdgeBank~\cite{poursafaei2022towards} shows that memorizing historical edges can be highly competitive on recurrence-heavy link-prediction benchmarks, indicating that future interactions often depend strongly on previously observed connectivity. Subsequent studies examine this redundancy from complementary perspectives: RepeatMixer~\cite{zou2024repeat} incorporates recurring higher-order interactions into neighborhood sampling, whereas TGB-Seq~\cite{yi2025tgbseq} reduces repeated edges to assess whether temporal models capture genuine sequential dynamics rather than recurrence shortcuts. These studies treat redundancy as an edge-level structural property, which is either exploited or removed for prediction.

However, structural recurrence alone does not reveal whether a node's state remains stable. A familiar neighborhood may carry a new conversation topic, while new neighbors may continue an existing one. Edge-level structural redundancy therefore cannot determine when a node state should be updated. HyperIntent instead measures redundancy at the level of intents and jointly considers structural and semantic novelty, so that sparse intent shifts, rather than dense edges, determine when node states are updated.

\section{Preliminaries}
\label{sec:preliminaries}

\subsection{Dynamic Textual Graphs}

We consider a dynamic textual graph as a chronologically ordered stream of directed interactions
\begin{equation}
\mathcal{E}=\{e_i=(u_i,v_i,t_i,r_i)\}_{i=1}^{N},
\end{equation}
where $e_i$ denotes the $i$-th interaction, $u_i,v_i$ are the source and destination nodes, $t_i$ is the timestamp, and $r_i$ is the interaction text. Multiple interactions from the same source node may share a timestamp. For a source node $u$ and timestamp $t$, we group these concurrent interactions into the source-specific batch
\begin{equation}
\mathcal{B}_{u,t}
=
\{e_i\in\mathcal{E}\mid u_i=u,\ t_i=t\}.
\end{equation}
To distinguish concurrent interactions at time $t$ from the events observed beforehand, we define the strict history at $t$ as
\begin{equation}
\mathcal{E}_{<t}=\{e_i\in\mathcal{E}\mid t_i<t\},
\qquad
\mathcal{G}_{<t}=(\mathcal{V},\mathcal{E}_{<t}),
\end{equation}
where $\mathcal{V}$ denotes the node set of the dynamic textual graph.
The historical neighbors of $u$ in $\mathcal{G}_{<t}$ are denoted by $\mathcal{N}_{u}(t)$. Every representation used to score an interaction at $t$ depends only on $\mathcal{E}_{<t}$. Events observed at $t$ can therefore update the semantic state only for subsequent predictions.

\subsection{Problem Definition}

Given the historical graph $\mathcal{G}_{<t}$, a dynamic textual graph model computes a time-dependent representation of node $u$ as
\begin{equation}
\mathbf{Z}_{u,t}=f_{\theta}(u|\mathcal{G}_{<t}),
\end{equation}
where $f_{\theta}$ is a parameterized encoder of the interactions observed before $t$, including their temporal and textual information. For a candidate interaction $(u,v,t)$, a task-specific decoder $g_{\phi}$ produces the prediction score 
\begin{equation}
\hat{y}_{u,v,t}
=
g_{\phi}(\mathbf{Z}_{u,t},\mathbf{Z}_{v,t}).
\end{equation}
We study two tasks using this score. For \emph{destination node retrieval}, the model receives a source node $u$, a query time $t$, and a candidate set $\mathcal{C}_{u,t}$. It ranks each $v\in\mathcal{C}_{u,t}$ by $\hat{y}_{u,v,t}$. For \emph{future link prediction}, the model predicts whether a candidate pair $(u,v)$ interacts at $t$. In both tasks, predictions at $t$ may depend only on $\mathcal{G}_{<t}$, thereby excluding concurrent and future interactions. Evaluation protocols are described in Section~\ref{sec:experiments} and Appendix~\ref{app:detailed-settings}.

\section{Methodology}

We present HyperIntent, a hypergraph-based framework that turns a dense textual edge stream into a sparse evolution of temporal intent. Given the dynamic textual graph defined in Section~\ref{sec:preliminaries}, HyperIntent constructs a node states cache by parsing continuously arriving edges into differentiated intent hyperedges, quantifying each hyperedge's temporal discrepancy against its semantic and structural histories, and performing shift-triggered updates only when intent shifts occur. During downstream prediction, HyperIntent retrieves the cached node states and fuses them with temporal structural representations produced by the TGNN backbone. Figure~\ref{fig:hyperintent-overview} illustrates the overall framework.

\subsection{Differentiated Intent Parsing}
\label{sec:parsing}

Continuously arriving edges often exhibit structural redundancy, and those associated with the same source node at a given timestamp may jointly express several distinct intents. Treating these interactions as isolated edges fragments each collective intent across multiple source–destination relations. Thus, HyperIntent maps each source-specific batch $\mathcal{B}_{u,t}$ to temporal intent hyperedges.

Within each batch, HyperIntent maps every relation text $r$ into a semantic space via $\phi_{\mathrm{lite}}(r)$ and groups semantically similar interactions into intent hyperedges. The resulting $k$-th intent hyperedge is defined as
\begin{equation}
e_k^H=(u,t,\mathcal{V}_k,\mathcal{R}_k),
\end{equation}
where $\mathcal{V}_k$ contains its incident destination nodes and $\mathcal{R}_k$ contains the relation texts of its member interactions. $e_k^H$ connects $u$ to $\mathcal{V}_k$ under their shared intent at $t$.

The relation-embedding multiset of the hyperedge is
\begin{equation}
\mathcal{X}_k=
\left\{\!\left\{
\phi_{\mathrm{lite}}(r)\mid r\in\mathcal{R}_k
\right\}\!\right\}.
\end{equation}
Together, $\mathcal{V}_k$ and $\mathcal{X}_k$ describe the structural incidence and textual semantics of the intent hyperedge, respectively.

Grouping intent-consistent interactions into separate hyperedges preserves concurrent intents while reducing redundancy among overlapping source--destination pairs. Each hyperedge thus serves as a sparse unit for subsequent discrepancy estimation and LLM reasoning.

\subsection{Temporal Intent Discrepancy Estimation}
\label{sec:discrepancy}

Many parsed hyperedges merely repeat previously observed intents rather than reflect intent shifts.  HyperIntent therefore evaluates each candidate with two complementary detectors in parallel: a semantic one compares its textual evidence with the node's semantic memory, and a structural one measures its deviation from historical neighborhoods. Their outputs jointly decide whether it is a shifted intent that triggers LLM reasoning.

\paragraph{\textbf{Semantic novelty detector.}}
For each node $u$, we maintain a persistent semantic memory
$\mathcal{M}_u^{<t}$, defined as the multiset union of relation embeddings
from all intent hyperedges identified as shifted strictly before $t$. The
memory is initialized as empty before the first shifted intent is observed.
It is used only for semantic novelty detection and is distinct from the
LLM-derived node states stored in the node states cache.

Given a candidate intent hyperedge $e_k^H$, the detector compares every relation embedding in $\mathcal{X}_k$ with its most similar element in $\mathcal{M}_u^{<t}$. The semantic discrepancy is defined as
\begin{equation}
d_{\mathrm{sem}}(e_k^H)=
\begin{cases}
1,
& \mathcal{M}_u^{<t}=\varnothing, \\[6pt]
\min\left\{1,\;
1-\dfrac{1}{|\mathcal{X}_k|}
\displaystyle\sum_{\mathbf{x}\in\mathcal{X}_k}
\max_{\mathbf{m}\in\mathcal{M}_u^{<t}}
\cos(\mathbf{x},\mathbf{m})\right\},
& \text{otherwise}.
\end{cases}
\end{equation}
A larger $d_{\mathrm{sem}}(e_k^H)$ indicates a greater departure from previously observed intent patterns. Since cosine similarity can be negative, the outer minimum bounds $d_{\mathrm{sem}}(e_k^H)$ within $[0,1]$, treating hyperedges that are on average no closer to the memory than orthogonal embeddings as maximally novel. When $\mathcal{M}_u^{<t}$ is empty, $e_k^H$ likewise receives the maximum discrepancy of 1, ensuring that the first intent of each node triggers LLM reasoning.

\paragraph{\textbf{Structural novelty detector.}}
We assess whether the member interactions of a candidate hyperedge depart from historical neighborhood patterns. For each member interaction $(u,v,t)$, we compute the temporal topological overlap between the historical two-hop neighborhoods $\mathcal{N}^{2}_u(t)$ and $\mathcal{N}^{2}_v(t)$ (nodes within two hops in $\mathcal{G}_{<t}$):
\begin{equation}
\operatorname{overlap}(u,v,t)=
\frac{
\left|\mathcal{N}^{2}_u(t)\cap\mathcal{N}^{2}_v(t)\right|
}{
\left|\mathcal{N}^{2}_u(t)\cup\mathcal{N}^{2}_v(t)\right|
},
\end{equation}
where the overlap is set to zero when the union is empty; two hops keep it well defined on bipartite graphs, where one-hop sets never intersect. A larger overlap indicates greater consistency with the historical neighborhood structure. The structural discrepancy of $e_k^H$ is defined as
\begin{equation}
d_{\mathrm{str}}(e_k^H)=
1-\frac{1}{|\mathcal{V}_k|}
\sum_{v\in\mathcal{V}_k}
\operatorname{overlap}(u,v,t).
\end{equation}
A larger $d_{\mathrm{str}}(e_k^H)$ indicates that the candidate hyperedge introduces a less familiar structural interaction pattern.

\paragraph{\textbf{Intent-shift decision.}}
The two novelty detectors capture complementary manifestations of temporal intent evolution: textual meaning may change while the interaction structure remains familiar, whereas new structural relations may emerge without an obvious semantic change. HyperIntent combines their decisions through an OR gate:
\begin{equation}
\delta_k=
\mathbf{1}
\left[
d_{\mathrm{sem}}(e_k^H)>\tau_{\mathrm{sem}}
\;\lor\;
d_{\mathrm{str}}(e_k^H)>\tau_{\mathrm{str}}
\right].
\end{equation}
Here, $\delta_k=1$ identifies $e_k^H$ as a shifted intent and selects it for LLM reasoning, whereas $\delta_k=0$ indicates that the existing node state can be inherited. 

After all intent hyperedges parsed from $\mathcal{B}_{u,t}$ have been
evaluated against the same pre-update memory $\mathcal{M}_u^{<t}$, the
semantic memory is updated as
\begin{equation}
\mathcal{M}_u^{t}
=
\operatorname{Update}
\left(
\mathcal{M}_u^{<t},
\left\{
\mathcal{X}_k \mid \delta_k=1
\right\}
\right).
\end{equation}
Here, $\mathrm{Update}(\cdot)$ appends the relation embeddings of all shifted intent hyperedges to the memory, leaving it unchanged if no $\delta_k=1$. The memory thus grows only with intent shifts and records which intents $u$ has expressed, while the recurrence of familiar intents is captured by the temporal encodings of the TGNN backbone. 

This parallel design reflects the central insight of HyperIntent: dense edge arrivals often correspond to sparse intent evolution. By invoking the LLM only when semantic or structural novelty signals a shift, HyperIntent captures complementary temporal changes without redundant semantic updates for stable interactions.

\subsection{Intent Representation Update}
\label{sec:update}

\begin{table*}[t]
\centering
\caption{Inductive future link prediction (AP, \%). Best and second-best results are in bold and underlined. Average rank ($\downarrow$) is over eight datasets (four for CROSS). }
\vspace{-2mm}
\label{tab:ap_auc_results}
\fontsize{6.8pt}{8.4pt}\selectfont
\renewcommand{\arraystretch}{1.08}
\setlength{\tabcolsep}{2pt}
\begin{tabularx}{1\textwidth}{@{}l l *{8}{>{\centering\arraybackslash}X} c@{}}
\toprule
\multicolumn{2}{c}{\multirow{2}{*}{Method}} & \multicolumn{8}{c}{Inductive AP} & \multirow{2}{*}{Avg. rank $\downarrow$} \\
\cmidrule(lr){3-10}
& & Enron & \mbox{ICEWS1819} & Googlemap & GDELT & \mbox{Stack elec} & \mbox{Stack ubuntu} & Amazon & Yelp & \\
\midrule
 & JODIE & \score{87.61}{0.23} & \score{93.33}{0.26} & OOM & \score{90.19}{0.23} & OOM & OOM & OOM & OOM & 12.62 \\
 & DyRep & \score{77.34}{0.44} & \score{91.34}{0.41} & OOM & \score{89.28}{0.11} & OOM & OOM & OOM & OOM & 13.62 \\
 & CAWN & \score{92.23}{0.11} & \score{96.31}{0.34} & \score{80.12}{0.21} & \score{89.86}{0.77} & \score{79.21}{0.86} & \score{78.82}{0.15} & \score{85.08}{0.06} & \score{90.10}{0.09} & 9.62 \\
 & TCL & \score{85.60}{0.24} & \score{97.89}{0.22} & \score{79.36}{0.09} & \score{91.51}{0.45} & \score{75.99}{0.41} & \score{77.77}{0.15} & \score{87.38}{0.50} & \score{92.49}{0.13} & 9.00 \\
 & GraphMixer & \score{83.28}{0.34} & \score{96.25}{0.30} & \score{76.33}{0.13} & \score{89.25}{0.48} & \score{81.42}{0.21} & \score{78.70}{0.15} & \score{85.17}{0.07} & \score{84.94}{0.08} & 11.12 \\
\cmidrule(l){1-11}
 & MoMent & \score{88.74}{0.10} & \score{96.63}{0.03} & \score{75.96}{0.05} & \score{86.36}{0.07} & \score{79.25}{0.06} & \score{80.97}{0.08} & \score{86.07}{0.08} & \score{87.16}{0.08} & 10.38 \\
 & PRISM & \score{94.13}{0.35} & \score{97.94}{0.05} & \score{85.25}{0.07} & \score{90.62}{0.03} & \score{84.61}{0.58} & \score{83.89}{0.68} & \score{88.01}{0.23} & \score{93.41}{0.08} & 5.88 \\
\cmidrule(l){1-11}
\multirow{4}{*}{\rotatebox[origin=c]{90}{TGAT}} & TGAT only & \score{85.89}{0.31} & \score{97.16}{0.33} & \score{87.50}{0.15} & \score{90.23}{0.10} & \score{83.91}{0.36} & \score{76.64}{0.15} & \score{87.60}{0.10} & \score{91.74}{0.10} & 8.50 \\
 & CROSS & \score{96.87}{0.47} & \secondscore{98.92}{0.01} & \bestscore{97.32}{0.04} & \score{87.24}{1.06} & - & - & - & - & 5.38 \\
 & DyGRASP & \score{91.66}{3.26} & \score{97.92}{0.02} & \score{95.31}{3.01} & \bestscore{93.05}{4.43} & \score{88.11}{0.31} & \score{82.67}{0.84} & \score{91.43}{0.29} & \score{93.85}{0.61} & 4.88 \\
\rowcolor{gray!12}
\cellcolor{white} & \textbf{HyperIntent} & \bestscore{98.16}{0.10} & \score{98.15}{0.06} & \score{96.01}{0.08} & \score{91.98}{0.01} & \score{91.58}{0.73} & \secondscore{87.92}{0.46} & \secondscore{93.41}{0.88} & \bestscore{95.37}{0.15} & \underline{2.88} \\
\cmidrule(l){1-11}
 & DyGFormer only & \score{94.09}{0.25} & \score{96.88}{0.18} & \score{77.35}{0.31} & \score{92.63}{0.09} & \score{88.01}{0.43} & \score{78.32}{0.15} & \score{87.80}{0.06} & \score{90.92}{0.06} & 7.38 \\
 & CROSS & \score{97.65}{1.20} & \score{98.91}{0.53} & \bestscore{97.32}{0.05} & \score{90.18}{0.59} & - & - & - & - & 4.38 \\
 & DyGRASP & \score{94.69}{0.72} & \score{98.20}{0.13} & \score{93.31}{0.11} & \score{92.11}{0.04} & \secondscore{92.50}{0.20} & \score{86.24}{0.92} & \score{92.76}{0.54} & \score{93.95}{0.27} & 3.75 \\
\rowcolor{gray!12}
\cellcolor{white}\multirow{-4}{*}{\rotatebox[origin=c]{90}{DyGFormer}} & \textbf{HyperIntent} & \secondscore{97.81}{0.21} & \bestscore{98.96}{0.06} & \secondscore{96.87}{0.31} & \secondscore{92.95}{0.16} & \bestscore{93.63}{0.81} & \bestscore{91.37}{0.65} & \bestscore{94.82}{0.33} & \secondscore{94.37}{0.15} & \textbf{1.62} \\
\addlinespace[2pt]
\bottomrule
\end{tabularx}
\end{table*}

After temporal discrepancy estimation, HyperIntent updates the textual state of a node only when at least one of its intent hyperedges exhibits a structural or semantic shift. Simply concatenating raw edge texts as input to an LLM can obscure the coherent interaction intent underlying fragmented edge-level records. We therefore assign each selected hyperedge to two role-specific agents that separately interpret its structural and semantic evidence.

For each selected intent hyperedge
$e_k^H=(u,t,\mathcal{V}_k,\mathcal{R}_k)$ with $\delta_k=1$,
the \emph{structural-role agent} examines the member group
$\mathcal{V}_k$ together with its current topological context and summarizes
the structural role represented by the hyperedge. In parallel, the
\emph{semantic-theme agent} analyzes the relation texts in $\mathcal{R}_k$
and summarizes their shared semantic theme. Denoting their outputs by
$s_k^{\mathrm{str}}$ and $s_k^{\mathrm{sem}}$, respectively, we construct
the structured hyperedge description as
\begin{equation}
S_k =
\operatorname{Compose}
\left(
s_k^{\mathrm{str}},
s_k^{\mathrm{sem}}
\right),
\end{equation}
where $\operatorname{Compose}(\cdot)$ deterministically organizes the two outputs into role-specific fields, preserving their structural and semantic information in a unified description. The task-specific agent prompts are provided in Appendix~\ref{app:prompts}.

The resulting description is embedded and normalized to obtain the
hyperedge-level representation:
\begin{equation}
\mathbf{E}_k^H
=
\operatorname{Normalize}
\left(
E_{\mathrm{text}}(S_k)
\right).
\end{equation}

HyperIntent aggregates the representations of all triggered intent hyperedges parsed from $\mathcal{B}_{u,t}$ to construct the semantic state of node $u$ at timestamp $t$. The cached semantic representation is updated as
\begin{equation}
\mathbf{E}_{u,t}
=
\operatorname{Mean}
\left(
\left\{
\mathbf{E}_k^H \mid \delta_k=1
\right\}
\right).
\end{equation}

Here, the set ranges over the intent hyperedges parsed from $\mathcal{B}_{u,t}$, and $\operatorname{Mean}(\cdot)$ denotes element-wise mean pooling. The record $(u,t,\mathbf{E}_{u,t})$ is appended to the timestamp-sorted node states cache only when at least one hyperedge satisfies $\delta_k=1$. Otherwise, the cache remains unchanged and no record is added.

\subsection{HyperIntent Training}

Given a query interaction $(u,v,t)$, HyperIntent constructs the node
representations using only the strict history $\mathcal{G}_{<t}$. The TGNN
backbone first encodes the historical interactions of node $u$ into the
temporal structural representation $\mathbf{H}_{u,t}$. 
In parallel, HyperIntent retrieves all states of node $u$ cached before query
time $t$. Let $\mathcal{T}_u^{<t}$ denote their update timestamps. HyperIntent averages the exponentially decayed cached states as
\begin{equation}
\widetilde{\mathbf{E}}_{u,t}
=
\operatorname{Mean}
\left(
\left\{
\exp\left(-\lambda(t-\tau)\right)\mathbf{E}_{u,\tau}
\mid
\tau\in\mathcal{T}_u^{<t}
\right\}
\right),
\end{equation}
where $\lambda\geq0$ controls the decay rate. This weighting preserves all cached intent shifts while downweighting those farther from query time $t$. Nodes without cached states use their encoded node text instead. Applying it to node $v$ yields $\widetilde{\mathbf{E}}_{v,t}$.

Then HyperIntent obtains the final node representation by applying an MLP to the concatenated temporal structural and decayed intent representations $\mathbf{Z}_{u,t}
=
\operatorname{MLP}
\left(
\mathbf{H}_{u,t}
\mathbin{\Vert}
\widetilde{\mathbf{E}}_{u,t}
\right),$
where $\mathbin{\Vert}$ denotes vector concatenation and
$\mathbf{Z}_{u,t}$ is the final representation of node $u$. The same
operations produce $\mathbf{Z}_{v,t}$ for the destination node $v$.

The resulting node representations are passed to the task-specific decoder $g_{\phi}$ defined in Section~\ref{sec:preliminaries} for interaction scoring. During training, the TGNN backbone, decay parameter $\lambda$, MLP, and task-specific decoder are optimized jointly, whereas the cached LLM-derived node states remain fixed. Consequently, LLM reasoning is confined to offline cache construction and does not participate in online training or inference.

\section{Experiments and Results}
\label{sec:experiments}

\begin{table*}[t]
\centering
\caption{Destination node retrieval (Hits@1, \%) under transductive and inductive settings. Best and second-best results are in bold and underlined. Average rank ($\downarrow$) is over 12 settings (8 for CROSS); OOM is ranked last and ``-‘’ is excluded.}
\vspace{-1mm}
\label{tab:hit1_results}
\fontsize{6.8pt}{8.4pt}\selectfont
\renewcommand{\arraystretch}{1.08}
\setlength{\tabcolsep}{1.5pt}
\begin{tabularx}{\textwidth}{@{}l l *{12}{>{\centering\arraybackslash}X} c@{}}
\toprule
\multicolumn{2}{c}{\multirow{2}{*}{Method}}
& \multicolumn{6}{c}{Transductive}
& \multicolumn{6}{c}{Inductive}
& \multirow{2}{*}{Avg. rank $\downarrow$} \\
\cmidrule(lr){3-8}\cmidrule(lr){9-14}
& & Enron & \mbox{ICEWS1819} & Googlemap & GDELT & Amazon & Yelp & Enron & \mbox{ICEWS1819} & Googlemap & GDELT & Amazon & Yelp & \\
\midrule
& JODIE
& \score{50.45}{0.83} & \score{66.03}{0.10} & OOM & \score{28.90}{0.31} & OOM & OOM
& \score{30.03}{0.42} & \score{51.34}{0.10} & OOM & \score{30.80}{0.74} & OOM & OOM & 12.25 \\

& DyRep
& OOM & \score{61.33}{0.26} & OOM & \score{31.12}{0.29} & OOM & OOM
& OOM & \score{47.99}{0.26} & OOM & \score{27.40}{0.51} & OOM & OOM & 13.50 \\

& CAWN
& \score{65.90}{0.18} & \score{78.12}{0.51} & \score{13.22}{0.19} & \score{42.53}{0.10} & \score{22.37}{0.26} & \score{32.18}{1.04} & \score{42.23}{0.29} & \score{63.42}{0.51} & \score{3.21}{0.04} & \score{29.16}{0.11} & \score{18.35}{0.42} & \score{26.42}{0.71} & 8.25 \\

& TCL
& \score{43.44}{0.20} & \score{81.88}{1.21} & \score{14.68}{0.04} & \score{43.53}{0.24} & \score{28.41}{0.13} & \score{15.54}{0.28} & \score{19.67}{0.14} & \score{60.59}{1.22} & \score{2.36}{0.05} & \score{31.99}{0.31} & \score{24.63}{0.08} & \score{13.83}{1.05} & 9.08 \\

& GraphMixer
& \score{44.05}{0.15} & \score{80.03}{0.78} & \score{12.02}{0.05} & \score{38.75}{0.33} & \score{26.13}{0.08} & \score{22.04}{0.37} & \score{20.46}{0.09} & \score{61.14}{0.78} & \score{1.91}{0.01} & \score{26.10}{0.46} & \score{22.24}{0.09} & \score{18.41}{0.14} & 10.75 \\

\midrule

& MoMent & \score{51.03}{0.35} & \score{75.94}{0.16} & \score{12.16}{0.02} & \score{43.10}{0.17} & \score{26.77}{0.97} & \score{27.16}{0.15} & \score{24.56}{0.25} & \score{54.35}{0.35} & \score{9.92}{0.04} & \score{17.66}{0.17} & \score{22.24}{0.98} & \score{22.50}{0.08} & 10.29 \\
& PRISM & \score{73.23}{0.81} & \score{82.59}{0.21} & \score{21.38}{0.05} & \score{50.49}{0.11} & \score{29.97}{0.54} & \secondscore{44.40}{0.43} & \score{49.70}{0.13} & \score{66.79}{0.49} & \score{17.46}{0.18} & \score{28.86}{0.25} & \score{25.23}{0.48} & \secondscore{37.03}{0.36} & 4.58 \\

\midrule

\multirow{4}{*}{\rotatebox[origin=c]{90}{TGAT}} & TGNN only
& \score{53.66}{0.26} & \score{78.09}{0.31} & \score{22.90}{0.42} & \score{41.35}{0.13} & \score{28.51}{0.09} & \score{37.37}{0.51} & \score{28.32}{0.33} & \score{57.52}{0.31} & \score{5.50}{0.07} & \score{26.30}{0.42} & \score{24.03}{0.38} & \score{30.86}{2.38} & 8.42 \\

& CROSS
& \score{48.18}{5.93} & \score{69.66}{1.29} & \score{20.65}{0.61} & \score{1.94}{1.91} & - & - & \score{29.28}{4.96} & \score{56.69}{0.63} & \score{18.46}{0.27} & \score{0.45}{0.31} & - & - & 11.12 \\

& DyGRASP
& \score{62.28}{0.07} & \score{80.26}{0.77} & \score{39.67}{0.50} & \score{46.82}{0.28} & \score{32.47}{0.81} & \score{43.81}{0.22} & \score{37.03}{0.09} & \score{61.14}{0.12} & \score{36.47}{0.29} & \score{24.39}{0.41} & \score{28.92}{0.85} & \score{35.87}{0.86} & 5.88 \\

\rowcolor{gray!12}
\cellcolor{white} & \textbf{HyperIntent}
& \score{73.92}{0.47} & \score{81.72}{0.12} & \bestscore{48.92}{0.14} & \score{48.61}{0.29} & \secondscore{37.49}{0.64} & \bestscore{45.93}{0.35} & \score{42.17}{0.51} & \score{63.11}{0.28} & \bestscore{42.94}{0.31} & \score{26.24}{0.27} & \secondscore{31.85}{0.44} & \bestscore{39.27}{0.13} & 3.67 \\

\midrule

& TGNN only
& \score{74.73}{0.23} & \score{80.36}{0.02} & \score{13.14}{0.02} & \score{47.64}{0.08} & \score{30.08}{0.07} & \score{35.12}{0.27} & \score{55.53}{0.27} & \score{63.40}{0.03} & \score{1.78}{0.06} & \bestscore{36.81}{0.38} & \score{26.13}{0.03} & \score{28.98}{0.36} & 6.00 \\

& CROSS
& \score{73.16}{2.94} & \score{72.85}{3.90} & \score{20.05}{0.80} & \score{18.47}{1.77} & - & - & \score{57.73}{2.45} & \score{61.19}{5.18} & \score{18.28}{0.44} & \score{7.00}{0.71} & - & - & 8.75 \\

& DyGRASP
& \secondscore{77.69}{2.91} & \secondscore{83.90}{0.36} & \secondscore{41.18}{0.42} & \secondscore{51.77}{0.02} & \score{34.76}{0.82} & \score{38.54}{0.15} & \secondscore{57.97}{3.74} & \secondscore{67.77}{0.76} & \secondscore{38.42}{0.50} & \score{30.79}{0.21} & \score{30.91}{0.33} & \score{33.74}{0.12} & \underline{2.92} \\

\rowcolor{gray!12}
\cellcolor{white}\multirow{-4}{*}{\rotatebox[origin=c]{90}{DyGFormer}} & \textbf{HyperIntent}
& \bestscore{81.65}{0.40} & \bestscore{87.21}{0.74} & \score{39.20}{0.71} & \bestscore{53.18}{0.27} & \bestscore{40.19}{0.43} & \score{43.68}{0.54} & \bestscore{61.28}{0.28} & \bestscore{73.48}{0.38} & \score{36.02}{0.54} & \secondscore{34.70}{0.24} & \bestscore{32.46}{0.37} & \score{36.23}{0.18} & \textbf{2.00} \\
\addlinespace[2pt]
\bottomrule
\end{tabularx}
\end{table*}

\begin{figure*}[!t]
    \centering
    \captionsetup{skip=2pt}
    \includegraphics[width=\textwidth]{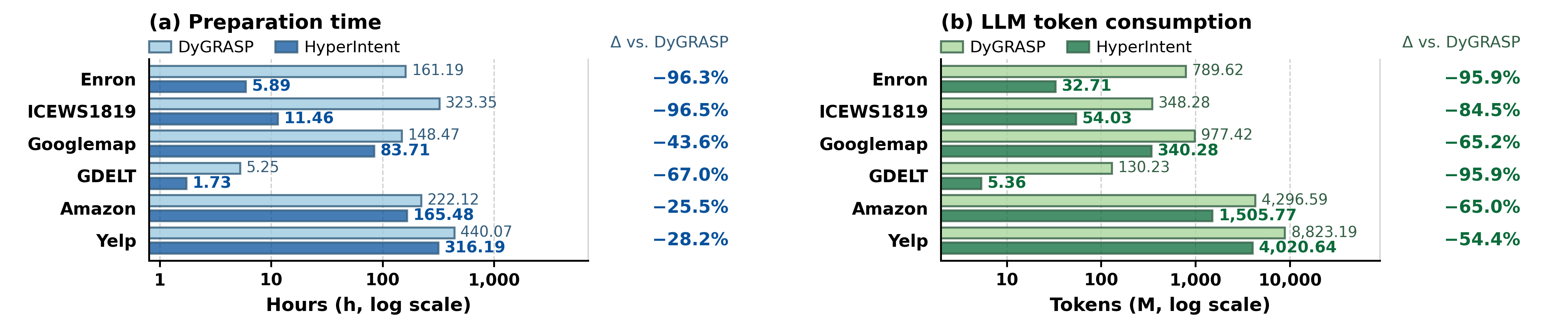}
    \caption{Efficiency comparison with DyGRASP: (a) preparation time and (b) LLM token consumption on logarithmic axes. Bar-end labels report absolute costs; the right-hand columns show HyperIntent's percentage change relative to DyGRASP.} 
    \label{fig:efficiency-comparison}
\end{figure*}

In this section, we evaluate the overall performance of HyperIntent by answering the following research questions (\textbf{RQ}s):
\begin{itemize}[leftmargin=1em]
    \item \textbf{RQ1:} Does HyperIntent improve predictive performance over baselines across TGNN backbones and evaluation settings?
    \item \textbf{RQ2:} How efficient is HyperIntent in terms of preparation time and LLM token consumption?
    \item \textbf{RQ3:} Which components of HyperIntent account for its performance gains?
    \item \textbf{RQ4:} When does HyperIntent update the cached intent states, and what do these states capture for destination retrieval?
    \item \textbf{RQ5:} How do the structural and semantic triggering thresholds affect predictive performance and encoding token consumption?
\end{itemize}

\subsection{Experimental Settings}

\paragraph{\textbf{Evaluation Tasks and Metrics.}}
Following the DTGB benchmark setting~\cite{zhang2024dtgb}, we evaluate under both transductive and inductive settings on two tasks: \textit{destination node retrieval}, measured by Hits@1, and \textit{future link prediction}, measured by average precision (AP) and AUC. The main comparison reports the performance of Hits@1 and inductive AP. Due to the limited space, the comprehensive results of transductive AP and transductive/inductive AUC results are reported in Appendix~\ref{app:additional-results}.

{\textbf{\textit{Datasets.}}}
We conduct experiments on eight datasets from the DTGB benchmark~\cite{zhang2024dtgb}, spanning diverse domains and interaction patterns: Enron, ICEWS1819, Googlemap, GDELT, Stack elec, Stack ubuntu, Amazon movies, and Yelp. Stack elec and Stack ubuntu support only future link prediction. Statistics and descriptions are provided in Appendix~\ref{app:detailed-settings}.

{\textbf{\textit{Baselines and Model Configurations.}}}
\begingroup
\setlength{\emergencystretch}{2em}
We compare HyperIntent with three groups of baselines. 
(1)~\emph{Temporal graph neural networks}, including JODIE~\cite{kumar2019jodie}, DyRep~\cite{trivedi2019dyrep}, CAWN~\cite{wang2021inductive}, TCL~\cite{wang2021tcl}, GraphMixer~\cite{cong2023really}, TGAT~\cite{xu2020inductive}, and DyGFormer~\cite{yu2023towards}. 
(2)~\emph{Modality-fusion methods}, including MoMent~\cite{xu2026unlocking} and PRISM~\cite{chang2026prism}, which encode text with a pretrained language model and fuse it with temporal and structural signals without invoking LLMs.
(3)~\emph{LLM-enhanced methods}, including CROSS~\cite{zhang2025cross}, DyGRASP~\cite{wang2026dygrasp}. Like HyperIntent, CROSS and DyGRASP enhance an existing TGNN backbone. For a fair comparison, we therefore implement all three methods on top of both TGAT and DyGFormer, ensuring that they share identical structural encoders. The remaining baselines define their own architectures and are evaluated in their original configurations. Results without public code are marked ``-''. Implementation details are given in Appendices~\ref{app:detailed-settings}.

\par\endgroup

\vspace{-1mm}
\subsection{Predictive Performance (RQ1)}

HyperIntent performs best overall on both tasks. Its DyGFormer-based variant achieves the lowest average rank on destination node retrieval (Table~\ref{tab:hit1_results}) and inductive future link prediction (Table~\ref{tab:ap_auc_results}), and its TGAT-based variant leads its backbone group on both tasks. Compared with the backbone-only models, HyperIntent improves mean Hits@1 by 34.0\% with TGAT and 25.4\% with DyGFormer. Under shared backbones, it also surpasses the state-of-the-art baseline DyGRASP in nearly all settings. The AP, AUC, and Hits@3/10 results in Appendices~\ref{app:additional-results} and~\ref{app:additional-retrieval} follow the same trend.

The gains are most pronounced when temporal structure is least informative. On Googlemap, the backbone-only models nearly fail on unseen nodes, reaching only 5.50 (TGAT) and 1.78 (DyGFormer) inductive Hits@1, whereas HyperIntent raises these scores to 42.94 and 36.02. In this regime, cached intent states provide semantic evidence that structural encoders cannot extract from sparse histories. In contrast, the gains are smaller on the event datasets GDELT and ICEWS1819. Their short, template-like relation texts and densely recurring interactions leave less room for intent-level semantics to complement temporal structure.

\subsection{Efficiency Analysis (RQ2)}
\label{sec:efficiency}

We compare HyperIntent with DyGRASP, the strongest LLM-enhanced baseline in RQ1, in terms of LLM token consumption and preparation time (Figure~\ref{fig:efficiency-comparison}). Both methods use the DyGFormer backbone and are measured under identical data splits, hardware, and LLM configurations (Appendix~\ref{app:encoding-savings}). 

\paragraph{\textbf{LLM token consumption.}} HyperIntent reduces total LLM token consumption by 54.4--95.9\% across all six datasets, accounting for both reasoning and representation encoding. 
The savings come from shift triggering: only shifted intent hyperedges invoke the LLM, while stable intents inherit their cached states. Counting representation encoding alone, HyperIntent saves 69\% on average relative to encoding every edge (Appendix~\ref{app:encoding-savings}).

\paragraph{\textbf{Preparation time.}} HyperIntent reduces preparation time by 43.6--96.5\% on Enron, ICEWS1819, Googlemap, and GDELT, but increases it by 159.6\% on Amazon and 61.6\% on Yelp. On these two largest datasets, the lightweight preprocessing stages (text encoding, intent parsing, and discrepancy estimation over large historical neighbor sets) dominate preparation time; these stages are counted in preparation time but not in LLM token consumption. This is a one-time cost: LLM reasoning is confined to offline cache construction and adds no overhead to training or inference.

\subsection{Ablation Study (RQ3)}
\label{sec:ablation}

To quantify the contribution of each module, we compare HyperIntent with five variants:
\begin{itemize}[leftmargin=1em]
    \item \textbf{\emph{w/o intent parsing}}: merges all concurrent edges of a source node into one hyperedge and caches it without shift selection.
    \item \textbf{\emph{w/o structural detection}} / \textbf{\emph{w/o semantic detection}}: triggers on semantic or structural novelty alone, i.e., $\delta_k=\mathbf{1}[d_{\mathrm{sem}}>\tau_{\mathrm{sem}}]$ or $\delta_k=\mathbf{1}[d_{\mathrm{str}}>\tau_{\mathrm{str}}]$.
    \item \textbf{\emph{w/o discrepancy}}: randomly selects as many intent hyperedges as HyperIntent triggers.
    \item \textbf{\emph{w/o shift-triggered updating}}: keeps intent parsing but caches every intent hyperedge without shift selection.
\end{itemize}

\begin{figure}[t]
    \centering
    \captionsetup{skip=2pt}
    \includegraphics[width=0.9\columnwidth]{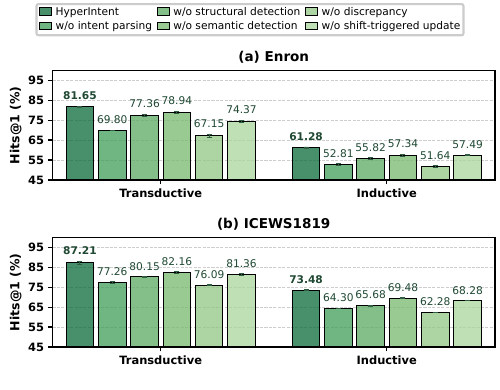}
    \caption{Ablation results on Enron and ICEWS1819 under the transductive and inductive settings. Performance is measured by Hits@1; error bars show the variation across runs.}
    \label{fig:rq3-ablation}
\end{figure}

As shown in Figure~\ref{fig:rq3-ablation}, HyperIntent outperforms all variants in every dataset--setting combination, and the average Hits@1 drops support three findings.
\textbf{(1) Which intents trigger updates matters most.} Random selection causes the largest drop (11.6 points), whereas caching every hyperedge costs 5.5 points. Selecting the same number of hyperedges at random is therefore worse than not selecting at all, so the gain comes from discrepancy-based selection rather than from caching fewer states.
\textbf{(2) Intent parsing is necessary.} Merging all concurrent edges into one hyperedge drops 9.9 points, 4.4 more than caching every parsed hyperedge, which isolates the contribution of separating concurrent intents.
\textbf{(3) The two detectors are complementary.} Removing structural or semantic detection drops 6.2 and 3.9 points, respectively. Structural novelty contributes more in every combination, yet the full model consistently outperforms the structure-only variant.

\subsection{Characterization of Intent Evolution (RQ4)}

We qualitatively examine how HyperIntent parses concurrent edges into intents, when it updates the cached states, and what the resulting intent representations preserve.

\begin{table}[!t]
\centering
\caption{Real concurrent interactions and the intent states inferred by HyperIntent. Multipliers indicate the numbers of concurrent edges summarized by each description.}
\label{tab:real-intent-examples}
\scriptsize
\setlength{\tabcolsep}{3pt}
\begin{tabularx}{\columnwidth}{@{}>{\raggedright\arraybackslash}X >{\raggedright\arraybackslash}X@{}}
\toprule
\textbf{Concurrent observed edges} & \textbf{Inferred intent states} \\
\midrule
\multicolumn{2}{@{}l@{}}{\textbf{(a) Enron \enspace|\enspace pete.davis@enron.com \enspace|\enspace $t=678$}} \\
\addlinespace[2pt]
\textbf{$\times 26$}\quad Hourly ISO final-schedule status reports: mostly no variances detected, with two generation-variance notices in the same status cluster.
&
\textbf{Schedule Status Monitoring.} Monitor hourly ISO schedule status, including routine completion and occasional generation variances.
\\
\addlinespace[2pt]
\textbf{$\times 14$}\quad Detailed hour-ahead download failures with SQL locks, empty-file diagnostics, or file-processing traces.
&
\textbf{Diagnostic Failure Recovery.} Diagnose repeated schedule-download failures using detailed SQL, file, and parsing evidence.
\\
\addlinespace[2pt]
\textbf{$\times 4$}\quad Brief schedule-download failure alerts stating that manual intervention is required.
&
\textbf{Manual Intervention Alert.} Escalate concise schedule-download failure notices that explicitly require manual intervention.
\\
\midrule
\multicolumn{2}{@{}l@{}}{\textbf{(b) Enron \enspace|\enspace rosalee.fleming@enron.com \enspace|\enspace $t=5$}} \\
\addlinespace[2pt]
\textbf{$\times 365$}\quad Calling All Alumni: invitation to a University of Houston reception hosted by Ken Lay and Chancellor Arthur K. Smith.
&
\textbf{University Alumni Outreach.} Coordinate a broad alumni invitation for an Enron and University of Houston reception.
\\
\addlinespace[2pt]
\textbf{$\times 1$}\quad Arrange immediate Enron Executive Database access for Mark LaGrange.
&
\textbf{Executive Access Provisioning.} Request immediate executive-database permissions for a specific employee.
\\
\addlinespace[2pt]
\textbf{$\times 1$}\quad Kenneth Lay approved Cindy Olson's attached expense report.
&
\textbf{Expense Approval Administration.} Transmit an approved employee expense report for administrative processing.
\\
\midrule
\multicolumn{2}{@{}l@{}}{\textbf{(c) ICEWS1819 \enspace|\enspace United States \enspace|\enspace $t=630$}} \\
\addlinespace[2pt]
\textbf{$\times 27$}\quad Host a visit involving Boris Johnson, Volodymyr Zelensky, Michel Aoun, heads of government, ministers, and other international leaders.
&
\textbf{Multilateral Diplomatic Hosting.} Coordinate a high-volume diplomatic reception involving national leaders and senior officials across regions.
\\
\bottomrule
\end{tabularx}
\vspace{-3mm}
\end{table}

\paragraph{\textbf{(1) Differentiated intent parsing.}}
Table~\ref{tab:real-intent-examples} pairs real concurrent edges with the intents inferred by HyperIntent. Parsing separates fine-grained variants of a shared topic: in case~(a), 44 schedule-related e-mails are divided into status monitoring, failure diagnosis, and manual-intervention alerts. It also prevents rare intents from being absorbed by dominant ones: in case~(b), two single-edge administrative requests remain distinct from a 365-edge alumni broadcast. Conversely, homogeneous activity collapses into a single intent, as with the 27 diplomatic interactions in case~(c).

\begin{figure}[!h]
    \centering
    \includegraphics[width=\columnwidth]{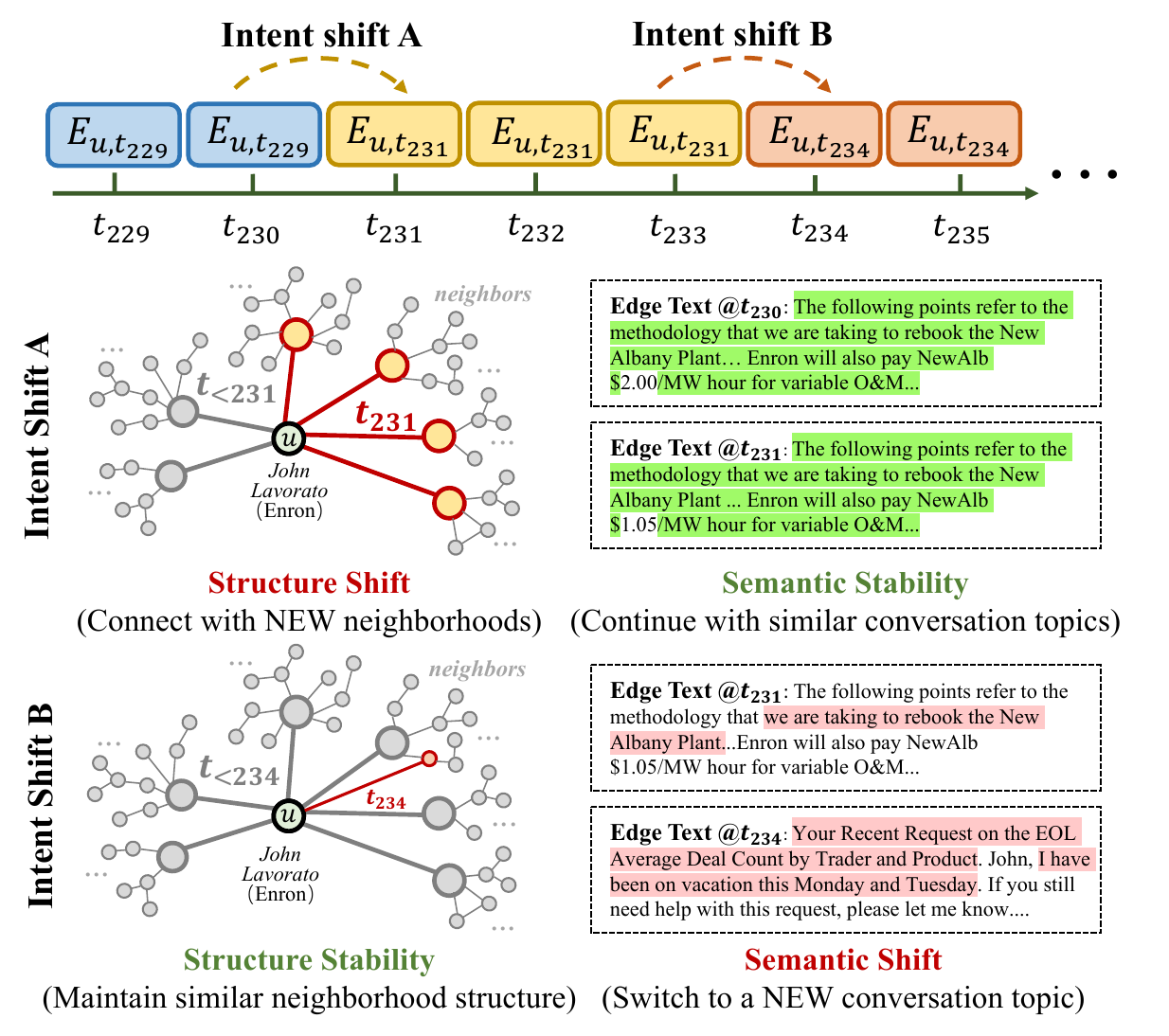}
    \caption{Shift-triggered state updates for an Enron node $u$ from $t_{229}$ to $t_{235}$. The cached state is inherited until an intent shift occurs, yielding only three distinct states over seven timestamps. Shift~A ($t_{231}$) is triggered by structural novelty under a stable topic, and Shift~B ($t_{234}$) by semantic novelty under a familiar neighborhood.}
    \vspace{-2mm}
    \label{fig:rq4-state-update}
\end{figure}

\paragraph{\textbf{(2) Shift-triggered state evolution.}}
Figure~\ref{fig:rq4-state-update} traces the cached states of an Enron node from $t_{229}$ to $t_{235}$. At $t_{231}$, the node connects to new neighborhoods while its edge text continues the New Albany plant topic from $t_{230}$; this structural shift triggers an update despite semantic stability. At $t_{234}$, the neighborhood remains familiar, but the conversation moves from plant rebooking to an EOL average deal-count request; this semantic shift triggers an update despite structural stability. Between shifts, the cached state is inherited, so seven timestamps of edge arrivals yield only three distinct states.

\begin{figure}[htbp]
    \centering
    \includegraphics[width=\columnwidth]{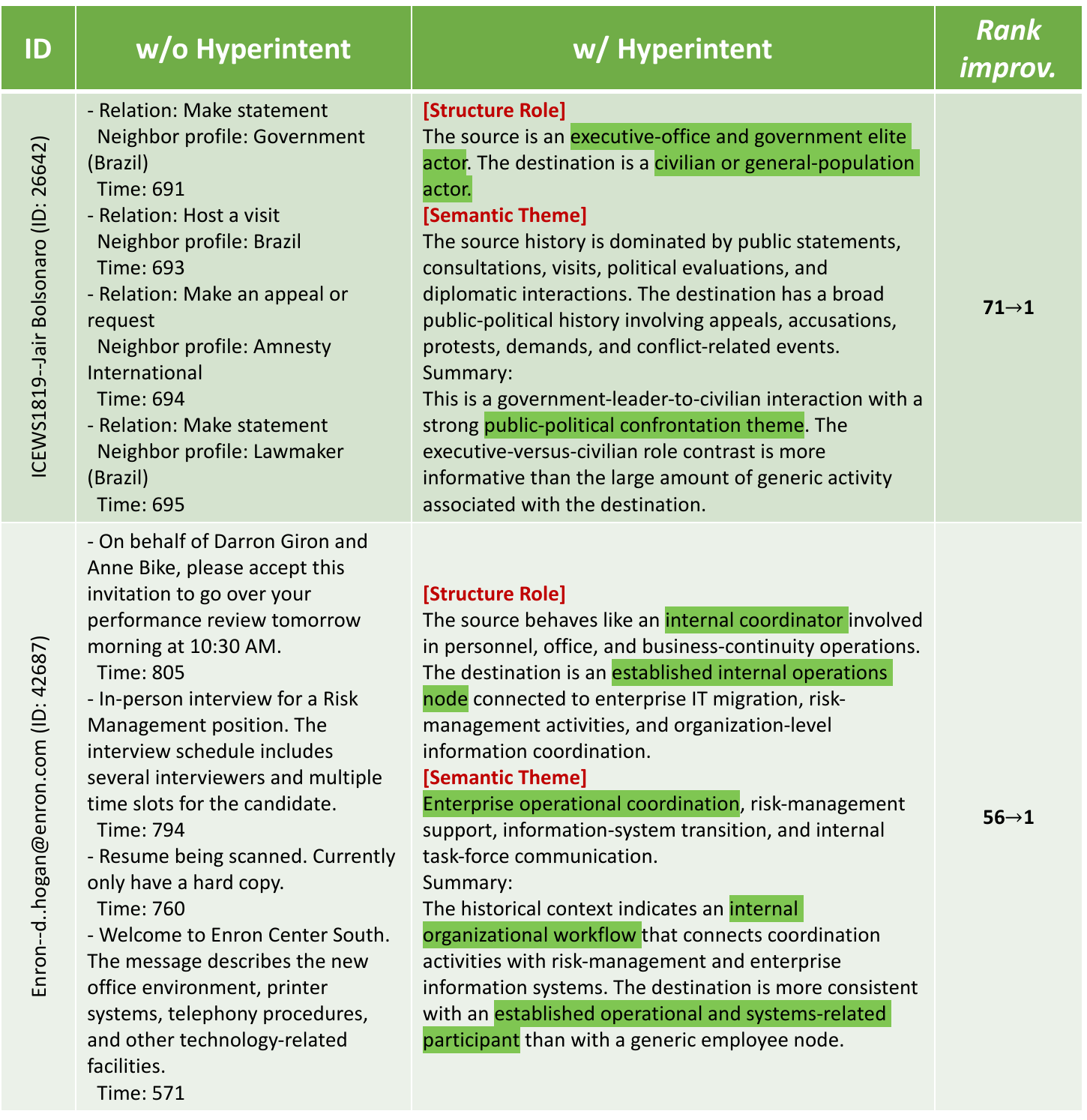}
    \caption{Destination-retrieval cases from ICEWS1819 and Enron. The left column shows the raw interaction history, and the right column shows the structured description generated by HyperIntent, where green highlights mark the role and theme cues supporting the correct target.}
    \label{fig:rq4-case-study}
\end{figure}

\paragraph{\textbf{(3) Intent representations for destination retrieval.}}
Figure~\ref{fig:rq4-case-study} shows how the resulting intent descriptions support retrieval. In ICEWS1819, the raw history mixes statements, visits, and requests involving heterogeneous political actors; HyperIntent identifies an executive-to-civilian role contrast and a public-political confrontation theme, raising the correct destination from rank 71 to rank 1. In Enron, the raw messages span performance reviews, recruitment, risk management, and information-system operations; HyperIntent consolidates them into an internal organizational workflow between a coordinator and an established operations node, raising the correct destination from rank 56 to rank 1.

Across all three stages, dense edge arrivals reduce to a small number of distinct intents and state transitions, consistent with the premise of HyperIntent.

\subsection{Triggering Threshold Sensitivity (RQ5)}
\label{sec:threshold-sensitivity}

\begin{figure}[t]
    \centering
    \includegraphics[width=\columnwidth]{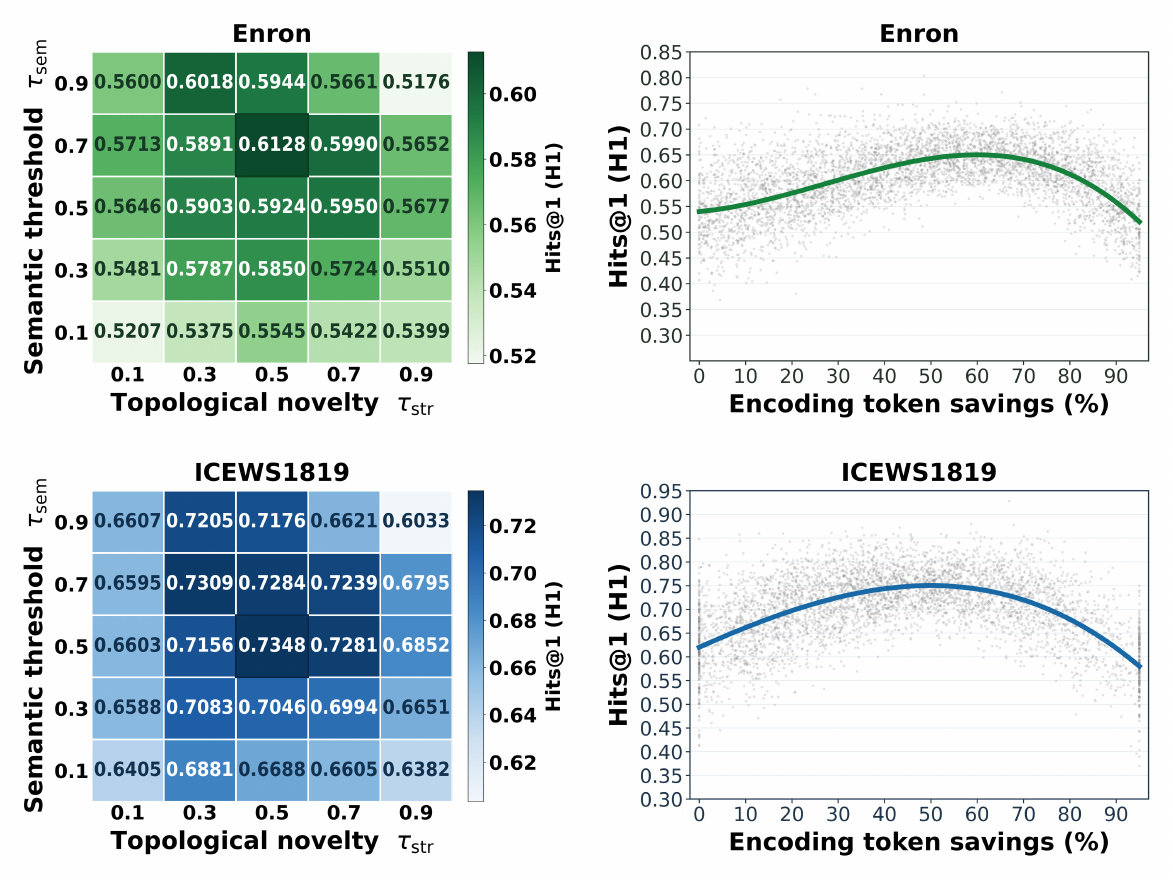}
    \caption{Effect of triggering thresholds on Enron (top) and ICEWS1819 (bottom). Heatmaps report Hits@1 over $(\tau_{\mathrm{str}},\tau_{\mathrm{sem}})$. In scatter plots, each gray point is one node batch under one threshold setting; curves are cubic fits.}
    \label{fig:rq5-sensitivity-efficiency}
    \vspace{-3mm}
\end{figure}

We examine how the structural and semantic triggering thresholds affect predictive performance and encoding token consumption. Using DyGFormer, we jointly vary $\tau_{\mathrm{str}}$ and $\tau_{\mathrm{sem}}$ on Enron and ICEWS1819 (Figure~\ref{fig:rq5-sensitivity-efficiency}). Hits@1 peaks at intermediate thresholds, reaching 0.6128 at $(\tau_{\mathrm{str}},\tau_{\mathrm{sem}})=(0.5,0.7)$ on Enron and 0.7348 at $(0.5,0.5)$ on ICEWS1819. Low thresholds trigger redundant semantic updates, whereas high thresholds miss informative intent shifts; both extremes degrade performance, so update frequency must be balanced against selective capture of temporal changes.

The batch-level fits show a corresponding pattern, with fitted Hits@1 peaking at approximately 60\% encoding token savings on Enron and 50\% on ICEWS1819. Beyond these levels, greater savings coincide with lower fitted Hits@1. These trends support the rationale for shift-triggered semantic updating: avoiding redundant updates while capturing informative intent shifts.

\section{Conclusion}

This paper introduced HyperIntent to model the sparse evolution of temporal intent beneath dense streams of edge arrivals in dynamic text-attributed graphs. Rather than treating every arriving edge as requiring a new semantic update, HyperIntent organizes concurrent interactions into differentiated intent hyperedges and updates node states only when structural or semantic discrepancies indicate a meaningful intent shift. Evaluations across multiple datasets, tasks, backbones, and evaluation settings show that this design improves predictive performance and LLM reasoning efficiency. More importantly, the ablations and temporal case studies explain these gains: intent parsing consolidates fragmented interactions, discrepancy estimation isolates meaningful temporal changes, and state inheritance preserves stable semantic context. Overall, these findings support shift-triggered intent evolution as an effective way to capture temporal changes while avoiding repeated encoding of stable interactions.

\clearpage

\bibliographystyle{ACM-Reference-Format}
\bibliography{references}

\clearpage
\appendix
\begingroup
\setlength{\emergencystretch}{2em}
\section{Experimental Details}
\label{app:detailed-settings}

\subsection{Datasets and Evaluation}
Table~\ref{tab:dataset-statistics} covers all eight DTGB datasets~\cite{zhang2024dtgb}. All contain node and edge text. Enron records employee e-mails; ICEWS1819 and GDELT record political events between named entities. Stack elec and Stack ubuntu contain multi-round dialogue interactions. Googlemap CT, Amazon movies, and Yelp contain user reviews of businesses or products. The two Stack datasets and the three review datasets are bipartite; Stack datasets are used only for link prediction.

\begin{table}[H]
\centering
\caption{Dataset statistics. Types denotes the number of edge categories; B indicates a bipartite graph. Googlemap CT is the Connecticut subset.}
\label{tab:dataset-statistics}
\small
\setlength{\tabcolsep}{3pt}
\renewcommand{\arraystretch}{1.05}
\begin{tabular*}{\columnwidth}{@{\extracolsep{\fill}}lrrrrc@{}}
\toprule
Dataset & Nodes & Edges & Times & Types & B \\
\midrule
Enron & 42,711 & 797,907 & 1,006 & 10 & No \\
ICEWS1819 & 31,796 & 1,100,071 & 730 & 266 & No \\
Googlemap CT & 111,168 & 1,380,623 & 55,521 & 5 & Yes \\
GDELT & 6,786 & 1,339,245 & 2,591 & 237 & No \\
Stack elec & 397,702 & 1,262,225 & 5,224 & 2 & Yes \\
Stack ubuntu & 674,248 & 1,497,006 & 4,972 & 2 & Yes \\
Amazon movies & 293,566 & 3,217,324 & 7,287 & 5 & Yes \\
Yelp & 2,138,242 & 6,990,189 & 6,036 & 5 & Yes \\
\bottomrule
\end{tabular*}
\end{table}

We use DTGB's chronological split (70\% training, 15\% validation, 15\% test), retrieval candidates, and link-prediction negative sampling. Transductive evaluation uses nodes observed in training; inductive evaluation tests unseen nodes. A query at $t$ accesses only $\mathcal{G}_{<t}$, excluding concurrent and future interactions. Our runs share the same splits and evaluation instances. Retrieval reports Hits@1/3/10; link prediction reports AP/AUC. Ablation, case, and sensitivity studies use retrieval. Each result table specifies its ranking scope; ties receive average ranks, missing results are excluded, and OOM receives the worst rank. 

\subsection{Implementation}
HyperIntent uses TGAT or DyGFormer with \texttt{all-\allowbreak MiniLM-\allowbreak L6-\allowbreak v2} (384 dimensions) for lightweight encoding and frozen \texttt{Llama-\allowbreak 3.1-\allowbreak 8B-\allowbreak Instruct} (4096 dimensions) for role-specific reasoning and cached representations. Intent grouping applies HDBSCAN~\cite{campello2013density} to the lightweight embeddings with cosine distance and \texttt{min\_cluster\_size}$=$2; edges labeled as noise are each kept as a singleton intent group. The cache is built once offline and remains fixed during training and inference; its triggering thresholds $(\tau_{\mathrm{str}},\tau_{\mathrm{sem}})$ are selected per dataset from $\{0.1,0.3,0.5,0.7,0.9\}^2$ by the validation score of task metric, and only the selected setting is evaluated on the test split. Trainable components use Adam, learning rate $10^{-4}$, batch size 256, and at most 100 epochs. Validation occurs every five epochs; early stopping uses five validation rounds, and testing uses the best validation checkpoint. Our experiments report mean and standard deviation over five independent seeds, using NVIDIA RTX 4090 and A100 GPUs.

\newpage
\subsection{Efficiency Measurement}
\label{app:efficiency-protocol}
{Figure~\ref{fig:efficiency-comparison}} measures complete offline preparation on six datasets with DyGFormer and matched chronological splits. HyperIntent includes lightweight encoding, intent parsing, discrepancy estimation, both reasoning calls, representation encoding, and cache storage. DyGRASP includes recent-context encoding, global reasoning, description encoding, and feature assembly. Timing starts with loaded models and empty caches and ends after feature storage. Measurements require matched hardware allocation, frozen LLM/tokenizer, precision, batch budget, and concurrency; each method retains its own contexts and prompts, with context and generation budgets recorded. LLM tokens include all non-padding input and generated output tokens for reasoning and representation encoding; reused text is counted on every processing pass. Lightweight encoding and detection count toward preparation time, but not LLM tokens.

\paragraph{Encoding-only savings.}
\label{app:encoding-savings}
Relative to sending every edge to the LLM, encoding-only savings are $(1-T_{\mathrm{HI}}^{\mathrm{enc}}/T_{\mathrm{full}}^{\mathrm{enc}})\times100\%$: Enron 66\%, ICEWS1819 51\%, Googlemap 76\%, and GDELT 83\% (unweighted mean 69\%). This baseline and token scope differ from the full DyGRASP preparation comparison.

\section{Prompt Templates}
\label{app:prompts}

The templates retain task-specific instructions and input fields; generic role initialization is omitted. Head node and member group denote $u$ and $\mathcal{V}_k$. Braced fields are populated with hyperedge evidence during offline cache construction.

\lstdefinestyle{prompt}{
    basicstyle=\ttfamily\scriptsize,
    breaklines=true,
    breakatwhitespace=true,
    columns=fullflexible,
    keepspaces=true,
    showstringspaces=false,
    frame=single,
    framerule=0.3pt,
    aboveskip=3pt,
    belowskip=3pt
}

\subsection{Structural-Role Prompt}

\begin{lstlisting}[style=prompt]
Summarize the structural interaction role of an intent-shift hyperedge
using only the supplied topological evidence.
Determine whether the member group:
- reinforces the head node's established neighborhood;
- extends the head node toward structurally unfamiliar neighbors; or
- combines familiar and unfamiliar interaction patterns.
Structural context:
- Time: {t}
- Number of destination nodes: {num_members}
- Mean historical topological overlap: {mean_overlap}
- Structural discrepancy: {d_str}
  (larger values indicate a less familiar structural interaction pattern)
Member-level structural evidence:
Member 1:
- Previously interacted with the head node: {yes | no}
- Historical topological overlap with the head node: {overlap_1}
Member 2:
- Previously interacted with the head node: {yes | no}
- Historical topological overlap with the head node: {overlap_2}
...
\end{lstlisting}

\subsection{Semantic-Theme Prompt}

\begin{lstlisting}[style=prompt]
Summarize the shared semantic theme and collective interaction intent of
the relation texts grouped into one temporal intent hyperedge. Treat them as a
group-level event rather than as independent pairwise interactions.
Relation texts:
1. {relation_text_1}
2. {relation_text_2}
...
N. {relation_text_n}
\end{lstlisting}

\par\endgroup

\onecolumn
\section{Supplementary Future Link Prediction Results}
\label{app:additional-results}
\begin{table}[H]
\centering
\renewcommand{\bestscore}[2]{\textbf{#1\textpm#2}}

\caption{Supplementary future link prediction results (transductive AP and transductive/inductive AUC, \%). Inductive AP is reported in Table~\ref{tab:ap_auc_results}. Best and second-best results are in bold and underlined. Average rank ($\downarrow$) is computed per block over eight datasets (four for CROSS), excluding unavailable results (-); ties share the average rank, and OOM receives the worst rank.}
\vspace{-2mm}
\label{tab:ap_auc_full_results}
\fontsize{7pt}{8.6pt}\selectfont
\renewcommand{\arraystretch}{1.02}
\setlength{\tabcolsep}{1.7pt}
\begin{tabularx}{\textwidth}{@{}l l *{8}{>{\centering\arraybackslash}X} c@{}}
\toprule
\multicolumn{2}{c}{Method} & Enron & \mbox{ICEWS1819} & Googlemap & GDELT & \shortstack{Stack\\elec} & \shortstack{Stack\\ubuntu} & \shortstack{Amazon\\movies} & Yelp & \shortstack{Avg. rank\\$\downarrow$} \\
\midrule
\multicolumn{11}{l}{\textbf{(a) Transductive AP}} \\
\midrule
 & JODIE & \score{95.53}{0.51} & \score{97.52}{0.37} & OOM & \score{94.66}{0.32} & OOM & OOM & OOM & OOM & 13.50 \\
 & DyRep & \score{90.66}{0.76} & \score{96.76}{0.26} & OOM & \score{94.16}{0.17} & OOM & OOM & OOM & OOM & 14.00 \\
 & CAWN & \score{97.56}{0.08} & \score{98.86}{0.25} & \score{87.21}{0.27} & \score{95.82}{0.53} & \score{95.29}{0.23} & \score{94.20}{0.05} & \score{88.85}{0.07} & \score{93.23}{0.27} & 9.62 \\
 & TCL & \score{96.03}{0.18} & \score{99.27}{0.12} & \score{83.35}{0.18} & \score{96.01}{0.11} & \score{94.41}{0.79} & \score{94.16}{0.23} & \score{90.51}{0.10} & \score{95.11}{0.13} & 9.12 \\
 & GraphMixer & \score{95.59}{0.27} & \score{98.71}{0.34} & \score{80.72}{0.10} & \score{95.23}{0.20} & \score{95.91}{0.09} & \score{94.16}{0.47} & \score{89.06}{0.06} & \score{88.83}{0.26} & 11.19 \\
\cmidrule(l){1-11}
 & MoMent & \score{96.73}{0.04} & \score{99.00}{0.00} & \score{80.89}{0.04} & \score{96.01}{0.03} & \score{95.21}{0.01} & \score{95.18}{0.02} & \score{89.84}{0.04} & \score{90.69}{0.07} & 9.56 \\
 & PRISM & \score{98.19}{0.09} & \score{99.30}{0.02} & \score{87.89}{0.14} & \score{96.99}{0.02} & \score{96.61}{0.13} & \score{95.98}{0.15} & \score{91.11}{0.17} & \score{95.75}{0.07} & 5.75 \\
\cmidrule(l){1-11}
\multirow{4}{*}{\rotatebox[origin=c]{90}{TGAT}} & TGAT only & \score{96.68}{0.26} & \score{99.08}{0.32} & \score{90.02}{0.19} & \score{95.72}{0.29} & \score{96.46}{0.05} & \score{93.52}{0.12} & \score{90.65}{0.16} & \score{94.57}{0.25} & 8.75 \\
 & CROSS & \bestscore{99.49}{0.20} & \secondscore{99.75}{0.01} & \bestscore{99.97}{0.00} & \score{94.63}{0.21} & - & - & - & - & 4.62 \\
 & DyGRASP & \score{98.04}{0.61} & \score{99.40}{0.02} & \score{96.23}{2.46} & \secondscore{97.78}{1.46} & \score{98.25}{0.13} & \score{95.38}{0.76} & \score{94.82}{0.51} & \score{97.12}{0.34} & 4.50 \\
\rowcolor{gray!12}
\cellcolor{white} & \textbf{HyperIntent} & \secondscore{99.39}{0.06} & \score{99.40}{0.04} & \score{96.93}{0.11} & \score{97.35}{0.04} & \secondscore{98.44}{0.29} & \secondscore{98.63}{0.85} & \secondscore{96.21}{0.43} & \bestscore{98.05}{0.67} & \underline{2.81} \\
\cmidrule(l){1-11}
 & DyGFormer only & \score{98.04}{0.15} & \score{99.01}{0.18} & \score{81.83}{0.38} & \score{96.53}{0.03} & \score{98.19}{0.10} & \score{94.31}{0.08} & \score{90.97}{0.03} & \score{93.91}{0.11} & 7.56 \\
 & CROSS & \score{99.28}{0.67} & \score{99.69}{0.23} & \bestscore{99.97}{0.00} & \score{96.23}{0.29} & - & - & - & - & 3.62 \\
 & DyGRASP & \score{98.69}{0.14} & \score{99.48}{0.04} & \score{94.55}{0.13} & \score{97.39}{0.02} & \score{98.42}{0.02} & \score{96.78}{0.92} & \score{95.34}{0.28} & \score{96.51}{0.73} & 3.88 \\
\rowcolor{gray!12}
\cellcolor{white}\multirow{-4}{*}{\rotatebox[origin=c]{90}{DyGFormer}} & \textbf{HyperIntent} & \score{99.17}{0.38} & \bestscore{99.88}{0.02} & \secondscore{97.43}{0.29} & \bestscore{98.08}{0.02} & \bestscore{99.12}{0.56} & \bestscore{98.89}{0.38} & \bestscore{96.75}{0.81} & \secondscore{97.24}{0.46} & \textbf{1.75} \\
\midrule
\multicolumn{11}{l}{\textbf{(b) Transductive AUC}} \\
\midrule
 & JODIE & \score{97.31}{0.52} & \score{97.41}{1.13} & OOM & \score{95.33}{0.20} & OOM & OOM & OOM & OOM & 13.00 \\
 & DyRep & \score{92.74}{0.26} & \score{96.32}{0.27} & OOM & \score{94.53}{0.18} & OOM & OOM & OOM & OOM & 14.00 \\
 & CAWN & \score{97.40}{0.07} & \score{98.57}{0.18} & \score{86.87}{0.63} & \score{96.00}{0.61} & \score{96.31}{0.07} & \score{95.67}{0.04} & \score{89.36}{0.12} & \score{93.49}{0.26} & 9.50 \\
 & TCL & \score{96.18}{0.25} & \score{99.23}{0.12} & \score{83.48}{0.94} & \score{96.19}{0.08} & \score{95.78}{1.06} & \score{95.16}{0.03} & \score{90.50}{0.12} & \score{95.28}{0.18} & 9.25 \\
 & GraphMixer & \score{95.67}{0.13} & \score{98.63}{0.24} & \score{80.95}{0.14} & \score{95.52}{0.18} & \score{96.73}{0.11} & \score{94.94}{0.28} & \score{88.94}{0.08} & \score{89.27}{0.21} & 11.38 \\
\cmidrule(l){1-11}
 & MoMent & \score{96.96}{0.04} & \score{98.97}{0.00} & \score{81.40}{0.06} & \score{96.21}{0.03} & \score{96.18}{0.02} & \score{95.92}{0.01} & \score{89.73}{0.04} & \score{90.88}{0.06} & 9.50 \\
 & PRISM & \score{98.13}{0.09} & \score{99.24}{0.02} & \score{88.00}{0.23} & \score{97.08}{0.02} & \score{97.22}{0.07} & \score{96.49}{0.07} & \score{91.12}{0.12} & \score{95.97}{0.06} & 5.88 \\
\cmidrule(l){1-11}
\multirow{4}{*}{\rotatebox[origin=c]{90}{TGAT}} & TGAT only & \score{96.81}{0.26} & \score{99.04}{0.39} & \score{90.49}{0.71} & \score{95.95}{0.33} & \score{97.09}{0.14} & \score{94.90}{0.18} & \score{90.64}{0.14} & \score{94.87}{0.29} & 8.88 \\
 & CROSS & \bestscore{99.43}{0.28} & \secondscore{99.73}{0.01} & \bestscore{99.97}{0.00} & \score{95.12}{0.17} & - & - & - & - & 4.62 \\
 & DyGRASP & \score{98.34}{0.39} & \score{99.43}{0.02} & \score{94.87}{0.04} & \score{97.22}{0.04} & \score{98.79}{0.06} & \score{96.38}{0.75} & \score{94.15}{0.53} & \score{97.62}{0.21} & 4.38 \\
\rowcolor{gray!12}
\cellcolor{white} & \textbf{HyperIntent} & \secondscore{99.38}{0.06} & \score{99.41}{0.03} & \score{97.17}{0.18} & \score{97.54}{0.04} & \score{98.44}{0.67} & \score{97.12}{0.89} & \secondscore{96.53}{0.34} & \secondscore{98.27}{0.58} & \underline{3.25} \\
\cmidrule(l){1-11}
 & DyGFormer only & \score{97.79}{0.14} & \score{98.88}{0.15} & \score{82.07}{0.18} & \score{96.62}{0.03} & \score{97.98}{0.06} & \score{95.26}{0.35} & \score{91.00}{0.06} & \score{94.07}{0.10} & 7.88 \\
 & CROSS & \score{99.26}{0.67} & \score{99.66}{0.26} & \bestscore{99.97}{0.00} & \score{96.40}{0.29} & - & - & - & - & 3.88 \\
 & DyGRASP & \score{98.77}{0.08} & \score{99.49}{0.04} & \score{94.64}{0.12} & \secondscore{97.59}{0.02} & \secondscore{98.87}{0.01} & \secondscore{97.84}{0.46} & \score{95.67}{0.82} & \score{96.31}{0.14} & 3.50 \\
\rowcolor{gray!12}
\cellcolor{white}\multirow{-4}{*}{\rotatebox[origin=c]{90}{DyGFormer}} & \textbf{HyperIntent} & \score{99.30}{0.56} & \bestscore{99.76}{0.02} & \secondscore{98.01}{0.34} & \bestscore{98.21}{0.02} & \bestscore{99.15}{0.73} & \bestscore{98.42}{0.35} & \bestscore{96.89}{0.51} & \bestscore{98.56}{0.68} & \textbf{1.50} \\
\midrule
\multicolumn{11}{l}{\textbf{(c) Inductive AUC}} \\
\midrule
 & JODIE & \score{87.32}{0.37} & \score{92.85}{0.65} & OOM & \score{89.21}{0.65} & OOM & OOM & OOM & OOM & 12.88 \\
 & DyRep & \score{79.01}{0.47} & \score{90.30}{0.97} & OOM & \score{89.17}{0.07} & OOM & OOM & OOM & OOM & 13.62 \\
 & CAWN & \score{90.91}{0.14} & \score{97.74}{0.39} & \score{70.58}{0.47} & \score{88.99}{0.82} & \score{79.63}{0.74} & \score{78.71}{0.80} & \score{84.92}{0.08} & \score{89.95}{0.05} & 9.88 \\
 & TCL & \score{85.12}{0.62} & \score{97.78}{0.12} & \score{78.95}{0.46} & \score{90.99}{0.22} & \score{76.89}{0.51} & \score{77.17}{0.27} & \score{86.77}{0.09} & \score{92.33}{0.11} & 8.69 \\
 & GraphMixer & \score{83.47}{0.39} & \score{96.05}{0.25} & \score{75.43}{0.18} & \score{89.42}{0.35} & \score{82.32}{0.31} & \score{78.70}{0.32} & \score{84.18}{0.12} & \score{84.52}{0.14} & 10.75 \\
\cmidrule(l){1-11}
 & MoMent & \score{88.52}{0.12} & \score{96.57}{0.02} & \score{75.22}{0.06} & \score{86.60}{0.11} & \score{79.22}{0.05} & \score{80.28}{0.07} & \score{85.14}{0.08} & \score{86.59}{0.10} & 10.38 \\
 & PRISM & \score{93.17}{0.51} & \score{97.78}{0.06} & \score{85.21}{0.12} & \score{90.31}{0.05} & \score{84.78}{0.36} & \score{83.29}{0.57} & \score{87.50}{0.18} & \score{93.29}{0.09} & 6.19 \\
\cmidrule(l){1-11}
\multirow{4}{*}{\rotatebox[origin=c]{90}{TGAT}} & TGAT only & \score{86.50}{0.32} & \score{97.06}{0.54} & \score{87.91}{0.28} & \score{90.12}{0.11} & \score{84.23}{0.18} & \score{76.55}{0.19} & \score{87.06}{0.23} & \score{91.73}{0.08} & 8.75 \\
 & CROSS & \score{96.79}{0.67} & \secondscore{98.79}{0.05} & \secondscore{97.23}{0.05} & \score{87.86}{0.93} & - & - & - & - & 5.50 \\
 & DyGRASP & \score{92.38}{2.33} & \score{97.96}{0.04} & \score{93.47}{0.08} & \score{90.74}{0.10} & \score{89.08}{0.18} & \score{82.71}{0.80} & \score{91.50}{0.32} & \score{93.78}{0.59} & 5.25 \\
\rowcolor{gray!12}
\cellcolor{white} & \textbf{HyperIntent} & \bestscore{98.12}{0.06} & \score{98.22}{0.05} & \score{96.34}{0.07} & \score{91.08}{0.06} & \score{91.49}{0.39} & \secondscore{87.88}{0.49} & \secondscore{93.38}{0.45} & \bestscore{95.30}{0.18} & \underline{2.62} \\
\cmidrule(l){1-11}
 & DyGFormer only & \score{93.16}{0.15} & \score{96.30}{0.27} & \score{76.48}{0.52} & \score{92.06}{0.03} & \score{86.07}{0.15} & \score{77.73}{0.47} & \score{87.33}{0.05} & \score{90.67}{0.09} & 7.50 \\
 & CROSS & \score{97.45}{1.22} & \score{98.77}{0.62} & \bestscore{97.30}{0.05} & \score{90.43}{0.59} & - & - & - & - & 3.50 \\
 & DyGRASP & \score{94.52}{0.59} & \score{98.13}{0.18} & \score{93.26}{0.10} & \secondscore{92.14}{0.03} & \secondscore{92.45}{0.22} & \score{86.30}{0.88} & \score{92.68}{0.50} & \score{94.01}{0.29} & 3.62 \\
\rowcolor{gray!12}
\cellcolor{white}\multirow{-4}{*}{\rotatebox[origin=c]{90}{DyGFormer}} & \textbf{HyperIntent} & \secondscore{97.63}{0.34} & \bestscore{99.20}{0.07} & \score{96.58}{0.39} & \bestscore{92.71}{0.17} & \bestscore{93.70}{0.77} & \bestscore{89.40}{0.68} & \bestscore{94.75}{0.35} & \secondscore{94.42}{0.18} & \textbf{1.50} \\
\bottomrule
\end{tabularx}
\end{table}

\clearpage
\section{Additional Destination Node Retrieval Results}
\label{app:additional-retrieval}
\begin{table}[H]
\centering
\renewcommand{\bestscore}[2]{\textbf{#1\textpm#2}}
\caption{Destination node retrieval (Hits@3, \%) under transductive and inductive settings. Best and second-best results are in bold and underlined. Average rank ($\downarrow$) is over 12 settings (8 for CROSS); OOM is ranked last and ``-'' is excluded.}
\vspace{-1mm}
\label{tab:hit3_results}
\fontsize{6.5pt}{8pt}\selectfont
\renewcommand{\arraystretch}{1.08}
\setlength{\tabcolsep}{1.5pt}
\begin{tabularx}{\textwidth}{@{}l l *{12}{>{\centering\arraybackslash}X} c@{}}
\toprule
\multicolumn{2}{c}{\multirow{2}{*}{Method}}
& \multicolumn{6}{c}{Transductive}
& \multicolumn{6}{c}{Inductive}
& \multirow{2}{*}{\shortstack{Avg. rank\\$\downarrow$}} \\
\cmidrule(lr){3-8}\cmidrule(lr){9-14}
& & Enron & \parbox[c]{\linewidth}{\centering ICEWS\\1819} & \parbox[c]{\linewidth}{\centering Googlemap\\CT} & GDELT & \parbox[c]{\linewidth}{\centering Amazon\\movies} & Yelp & Enron & \parbox[c]{\linewidth}{\centering ICEWS\\1819} & \parbox[c]{\linewidth}{\centering Googlemap\\CT} & GDELT & \parbox[c]{\linewidth}{\centering Amazon\\movies} & Yelp & \\
\midrule
& JODIE
& \score{76.12}{3.11} & \score{85.32}{0.12} & OOM & \score{57.52}{0.35} & OOM & OOM & \score{52.44}{0.16} & \score{71.40}{0.86} & OOM & \score{52.94}{0.28} & OOM & OOM & 11.92 \\

& DyRep
& OOM & \score{84.15}{0.61} & OOM & \score{57.31}{0.28} & OOM & OOM & OOM & \score{70.10}{0.32} & OOM & \score{51.79}{0.68} & OOM & OOM & 13.25 \\

& CAWN
& \score{77.11}{0.64} & \score{89.51}{0.78} & \score{27.71}{0.18} & \score{66.31}{0.17} & \score{41.22}{0.65} & \score{53.93}{0.24} & \score{59.62}{0.27} & \score{81.20}{0.66} & \score{8.28}{0.21} & \score{46.67}{0.61} & \score{34.41}{0.51} & \score{45.53}{1.20} & 9.17 \\

& TCL
& \score{70.08}{0.81} & \score{93.56}{1.06} & \score{29.66}{0.31} & \score{67.56}{0.29} & \score{48.00}{0.11} & \score{29.86}{0.84} & \score{39.28}{0.35} & \score{80.26}{0.91} & \score{6.21}{0.08} & \score{50.90}{0.24} & \score{41.57}{0.23} & \score{25.94}{1.68} & 9.00 \\

& GraphMixer
& \score{67.54}{0.27} & \score{92.31}{0.13} & \score{25.14}{0.28} & \score{63.51}{0.30} & \score{44.38}{0.13} & \score{40.17}{0.42} & \score{39.16}{0.11} & \score{80.01}{0.88} & \score{4.65}{0.04} & \score{47.55}{0.81} & \score{37.64}{0.11} & \score{34.13}{0.60} & 10.83 \\

\midrule

& MoMent & \score{74.86}{0.16} & \score{91.07}{0.02} & \score{25.58}{0.09} & \score{67.59}{0.24} & \score{45.65}{1.33} & \score{46.58}{0.01} & \score{45.16}{0.45} & \score{75.74}{0.22} & \score{20.46}{0.06} & \score{35.58}{0.24} & \score{38.42}{1.41} & \score{39.48}{0.01} & 10.17 \\
& PRISM & \score{85.97}{0.48} & \score{93.61}{0.14} & \score{39.65}{0.10} & \score{73.89}{0.06} & \score{50.19}{0.50} & \bestscore{66.39}{0.46} & \score{70.36}{0.20} & \score{84.13}{0.54} & \score{33.77}{0.22} & \score{47.79}{0.19} & \score{43.23}{0.57} & \bestscore{57.63}{0.46} & 4.33 \\

\midrule

\multirow{4}{*}{\rotatebox[origin=c]{90}{TGAT}} & TGNN only
& \score{75.83}{0.82} & \score{91.88}{0.34} & \score{43.25}{0.23} & \score{66.21}{0.19} & \score{48.55}{0.13} & \score{59.68}{0.43} & \score{50.45}{0.25} & \score{78.35}{0.29} & \score{15.45}{0.17} & \score{48.18}{0.55} & \score{41.65}{0.61} & \score{51.20}{2.06} & 8.17 \\

& CROSS
& \score{73.75}{4.32} & \score{86.63}{0.88} & \score{33.66}{1.77} & \score{7.16}{6.97} & - & - & \score{51.89}{4.21} & \score{78.64}{0.36} & \score{29.27}{0.74} & \score{1.58}{1.55} & - & - & 11.38 \\

& DyGRASP
& \score{84.49}{4.34} & \score{93.45}{0.17} & \score{60.91}{0.26} & \score{71.68}{0.16} & \score{51.23}{0.45} & \secondscore{63.78}{0.21} & \score{63.01}{6.71} & \score{81.69}{0.36} & \score{54.84}{0.43} & \score{43.72}{0.25} & \score{45.67}{0.34} & \score{54.23}{0.51} & 5.00 \\

\rowcolor{gray!12}
\cellcolor{white} & \textbf{HyperIntent}
& \score{86.42}{0.73} & \score{92.15}{0.48} & \bestscore{64.37}{0.89} & \score{71.84}{0.25} & \secondscore{55.62}{0.51} & \score{63.74}{0.36} & \score{61.27}{0.54} & \score{80.46}{0.81} & \score{55.18}{0.66} & \score{49.72}{0.39} & \secondscore{47.85}{0.75} & \secondscore{57.23}{0.28} & 4.00 \\

\midrule

& TGNN only
& \score{88.77}{0.51} & \score{91.75}{0.06} & \score{27.05}{0.21} & \score{70.89}{0.10} & \score{49.90}{0.08} & \score{55.49}{0.54} & \score{74.74}{0.26} & \score{80.17}{0.71} & \score{4.48}{0.07} & \bestscore{56.66}{0.26} & \score{43.19}{0.11} & \score{47.51}{0.64} & 6.83 \\

& CROSS
& \score{88.33}{3.28} & \score{89.41}{2.19} & \score{35.37}{1.49} & \score{42.33}{1.43} & - & - & \score{78.97}{2.83} & \score{80.20}{4.19} & \score{29.71}{2.94} & \score{16.71}{0.82} & - & - & 8.50 \\

& DyGRASP
& \secondscore{94.48}{1.01} & \secondscore{95.00}{0.24} & \secondscore{63.13}{0.59} & \secondscore{75.36}{0.07} & \score{53.45}{0.67} & \score{59.12}{0.88} & \secondscore{84.38}{1.19} & \secondscore{84.88}{0.65} & \bestscore{57.52}{0.71} & \score{49.54}{0.08} & \score{46.89}{0.72} & \score{52.14}{0.39} & \underline{3.08} \\

\rowcolor{gray!12}
\cellcolor{white}\multirow{-4}{*}{\rotatebox[origin=c]{90}{DyGFormer}} & \textbf{HyperIntent}
& \bestscore{95.18}{0.62} & \bestscore{95.41}{0.87} & \score{58.23}{0.14} & \bestscore{76.96}{0.53} & \bestscore{56.78}{0.92} & \score{62.45}{0.41} & \bestscore{86.34}{0.42} & \bestscore{86.59}{0.68} & \secondscore{56.67}{0.91} & \secondscore{53.28}{0.57} & \bestscore{48.12}{0.33} & \score{55.46}{0.84} & \textbf{1.83} \\
\addlinespace[2pt]
\bottomrule
\end{tabularx}
\end{table}

\begin{figure}[H]
    \centering
    \captionsetup{skip=2pt}
    \includegraphics[width=0.9\textwidth]{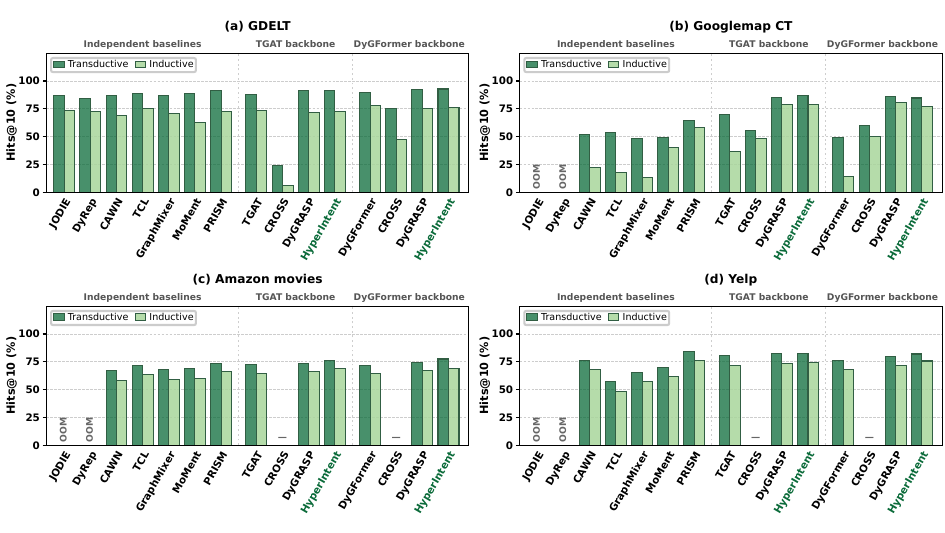}
    \caption{Destination node retrieval (Hits@10, \%) on (a) GDELT, (b) Googlemap CT, (c) Amazon movies, and (d) Yelp. Dark and light green bars denote the transductive and inductive settings, respectively. Methods are grouped as independent baselines, TGAT-based methods, and DyGFormer-based methods. OOM denotes out-of-memory failures; ``---'' indicates unavailable results.}
    \label{fig:hit10_results}
\end{figure}

\end{document}